\documentclass[manuscript]{aastex62}
\usepackage{amsbsy}
\usepackage{amsfonts}
\usepackage{amssymb}
\usepackage{amsmath}

\usepackage{booktabs}
\usepackage{array}

\graphicspath{{./}{figures/}}

\received{xxxx}
\revised{yyyy}
\accepted{zzzz}
\submitjournal{ApJ}

\shorttitle{Energy decrease of an Alfv{\'e}n wave passing through an Alfv{\'e}n speed gradient}
\shortauthors{Bose et al.}

\begin{document}

\title{Measured reduction in Alfv{\'e}n wave energy propagating through longitudinal gradients scaled to match solar coronal holes}

\author[0000-0001-8093-9322]{Sayak Bose}
\affiliation{Columbia Astrophysics Laboratory, Columbia University, 550 West 120th Street, New York, NY 10027 USA} \email{bose.sayak16@gmail.com}

\author[0000-0002-5741-0495]{Troy Carter}
\affiliation{Department of Physics and Astronomy, University of California, Los Angeles, California 90095, USA}

\author[0000-0001-7748-4179]{Michael Hahn}
\affiliation{Columbia Astrophysics Laboratory, Columbia University, 550 West 120th Street, New York, NY 10027 USA}

\author[0000-0002-6500-2272]{Shreekrishna Tripathi}
\affiliation{Department of Physics and Astronomy, University of California, Los Angeles, California 90095, USA}

\author[0000-0002-6468-5710]{Stephen Vincena}
\affiliation{Department of Physics and Astronomy, University of California, Los Angeles, California 90095, USA}

\author[0000-0002-1111-6610]{Daniel Wolf Savin}
\affiliation{Columbia Astrophysics Laboratory, Columbia University, 550 West 120th Street, New York, NY 10027 USA}



\begin{abstract}

We have explored the effectiveness of a longitudinal gradient in Alfv{\'e}n speed in reducing the energy of propagating Alfv{\'e}n waves under conditions scaled to match solar coronal holes. The experiments were conducted in the Large Plasma Device at the University of California, Los Angeles. Our results show that the energy of the transmitted Alfv{\'e}n wave decreases as the inhomogeneity parameter, $\lambda/L_{\rm A}$, increases. Here, $\lambda$ is the wavelength of the Alfv{\'e}n wave and $L_{\rm A}$ is the scale length of Alfv{\'e}n speed gradient. For gradients similar to those in coronal holes, the waves are observed to lose a factor of $\approx 5$ more energy than they do when propagating through a uniform plasma without a gradient. We have carried out further experiments and analyses to constrain the cause of wave energy reduction in the gradient. The loss of Alfv{\'e}n wave energy from mode coupling is unlikely, as we have not detected any other modes. Contrary to theoretical expectations, the reduction in the energy of the transmitted wave is not accompanied by a detectable reflected wave. Nonlinear effects are ruled out as the amplitude of the initial wave is too small and the wave frequency well below the ion cyclotron frequency. Since the total energy must be conserved, it is possible that the lost wave energy is being deposited in the plasma. Further studies are needed to explore where the energy is going.

\end{abstract}


\keywords{Sun: corona, magnetic fields, magnetohydrodynamics (MHD), plasmas, waves, (Sun:) solar wind }


\section{Introduction}\label{sec:intro}

Coronal holes are regions of the Sun's atmosphere with open magnetic field lines that extend into interplanetary space. These regions are $\sim$ 200 times hotter than the underlying photosphere.  It is widely established that the fast solar wind originates from coronal holes; but the mechanism responsible for heating coronal holes and accelerating the fast solar wind remains a mystery \citep{Cranmer2009}.  

Recent observations at the base of coronal holes have detected Alfv{\'e}nic waves with sufficient energy to heat coronal holes and accelerate the fast solar wind \citep{mcintosh2011alfvenic,morton2015investigating}. Furthermore, strong damping of Alfv{\'e}nic waves has been seen at a height of $\approx 0.15 \;\rm R_{\odot}$, where $\rm R_{\odot}$ is the solar radius, implying that coronal holes are predominantly heated by wave-driven processes \citep{bemporad2012spectroscopic, hahn2012evidence,hahn2013observational}. Here and throughout, all distances in coronal holes are measured from the surface of the Sun. The term Alfv{\'e}nic is used to highlight that some of the observed waves may not be pure torsional Alfv{\'e}n waves.  Transverse kink waves may also be present \citep{van2008detection, goossens2009nature, goossens2012surface}. Studies of chromospheric spicules by \cite{de2012ubiquitous} suggest that both modes contribute to the coronal wave energy.

Different models have been put forward to explain the damping of wave energy in coronal holes  \citep{moore1991alfven,Moore_1991,ofman1995alfven,hood1997heating, Matthaues1999, Dimtruk_2001, oughton_2001}. A number of these models invoke partial reflection of the upward propagating torsional Alfv{\'e}n waves \citep{moore1991alfven,Moore_1991, Matthaues1999, Dimtruk_2001, oughton_2001}. This wave reflection is thought to be caused by a strong longitudinal gradient in Alfv{\'e}n speed along the magnetic field lines at low heights in coronal holes \citep{moore1991alfven,musielak1992klein}.

Most of the experiments to date on the propagation of torsional Alfv{\'e}n waves through a longitudinal Alfv{\'e}n speed gradient were motivated by the needs of fusion devices, such as mirror machines. Alfv{\'e}n speed gradients were produced by introducing a nonuniformity in the magnetic field of the machine. Torsional Alfv{\'e}n waves were excited in the high magnetic field region. These waves propagated along the field lines into a region of decreasing magnetic field to the point where the wave frequency matched the local ion-cyclotron frequency, a configuration known as a magnetic beach, causing ion heating. The efficiency of this wave-driven heating was studied in mirror machines \citep{swanson1972rf,breun1987stabilization,roberts1989m} using different types of antennas \citep{Stix:58,yasaka1988icrf}. A few basic plasma physics experiments have also been carried out to study the characteristics of a torsional Alfv\'en wave through a longitudinal gradient \citep{vincena2001shear, mitchell2002laboratory}.  Propagation of Alfv\'en waves through gradients produced by periodically arranged multiple magnetic wells was studied by \citet{zhang2008spectral}. However, in all the above mentioned experiments, either the gradient was too weak, or the geometry of the gradient was different compared to coronal holes.

Here we report new laboratory experiments to study the propagation of torsional Alfv{\'e}n waves through a longitudinal Alfv{\'e}n speed gradient under conditions scaled to match those of coronal holes. The wave experiments were carried out in the Large Plasma Device (LAPD), located at the University of California, Los Angeles \citep{gekelman2016upgraded}.   

The rest of the paper is organized as follows: In Section \ref{corona_lab} we describe the basic physics of torsional Alfv{\'e}n waves and compare the  plasma conditions and properties of  Alfv{\'e}n waves in a coronal hole with those in LAPD. The experimental set up is described in Section \ref{EXp}. The results of our wave experiments are presented in Section \ref{Obs} and analyzed in Sections \ref{Sec:analysis}. This is followed by a discussion and summary in Section \ref{dis_sum}.

\section{Alfv{\'e}n waves}\label{corona_lab}

\subsection{Overview}

Alfv{\'e}n waves are one of the fundamental wave modes of magnetized plasmas. These waves were first predicted by \citet{alfven1942existence} using ideal magnetohydrodynamics (MHD). In cylindrical geometry they are commonly referred to as torsional Alfv\'en waves, while in Cartesian coordinates they are often referred to as shear Alfv\'en waves.  From hereon we will use the term shear Alfv{\'e}n waves. These waves cause shearing and twisting of magnetic field lines. The resulting magnetic tension provides the restoring force for the waves. 

Shear Alfv{\'e}n waves are low frequency electromagnetic waves that propagate below the ion cyclotron frequency, $\omega_{\rm ci}=qB_0/m_{\rm i}$, where $q$ is the ion charge, $B_0$ is the magnitude of the ambient magnetic field, and $m_{\rm i}$ is the ion mass. In the ideal MHD limit, these waves transport energy along the ambient magnetic field lines and follow the linear dispersion relation, 
\begin{equation}
\omega=v_{\rm A}k_{\parallel}. 
\label{Eq:Alfven_dispersion}
\end{equation} 
\noindent
Here, $\omega$ is the frequency of the wave in units of rad $\rm s^{-1}$, $v_{\rm A}$ is the Alfv{\'e}n speed, and $k_{\parallel}$ is the wave number parallel to the ambient magnetic field. The Alfv{\'e}n speed is given by $v_{\rm A}=B_{\rm 0}/\sqrt{\mu_{0}\rho}$,  where $\mu_{0}$ is the permeability of free space, $\rho=\left( n_{\rm i} m_{\rm i} + n_{\rm e} m_{\rm e} \right)$ is the mass density of the plasma, $n_{\rm i}$ is the ion number density, $n_{\rm e}$ is the electron number density, and $m_{\rm e}$ is the electron mass \citep{alfven1942existence, cross1988introduction, priest_2014}. For quasineutral plasmas, $n_{\rm i} \simeq n_{\rm e} \simeq n$, and $n$ is usually referred to as the plasma density.

Shear Alfv\'en waves interact with the plasma and drive ion and electron currents. Ideal MHD includes the perpendicular motion of the ions in the wave dynamics, but this theory does not explicitly describe the parallel response of electrons. This aspect of shear Alfv\'en wave dynamics is considered by more advanced theories, such as two-fluid theory, plasma kinetic theory, etc. A commonly used dimensionless parameter to describe the parallel response of an electron is $\bar{\beta}\equiv 2v_{\rm te}^2 /\left( \omega /k_{\parallel} \right)^2\approx~2~v_{\rm te}^2/v_{\rm A}^2$, where $v_{\rm te}=\sqrt{ T_{\rm e}/ m_{\rm e}}$ is the electron thermal velocity and $T_{\rm e}$ is the electron temperature. In this paper, $T_{\rm e}$ is expressed in joules in all the formulae unless stated otherwise.

For $\bar{\beta}\gg1$, the electrons respond adiabatically to the wave field and the wave is called a kinetic Alfv{\'e}n wave (KAW). The term KAW is also used by some authors for shear Alfv\'en waves influenced by the ion gyroradius, but we refer specifically to the $\bar{\beta}\gg1$ regime. The dispersion relation of a KAW is given by \citep{stasiewicz2000small}
\begin{equation}
    \frac{\omega}{k_{\parallel}}=v_{\rm A}\sqrt{1-\bar{\omega}^{2}\left(1+ k_{\perp}^2 \rho_{\rm i}^2\right)+k_{\perp}^2\left(\rho_{\rm s}^{2}+\rho_{\rm i}^2\right)},
    \label{KAW_dispersion}
\end{equation}
\noindent
where $\bar{\omega}$ $=\omega/\omega_{\rm ci}$, $\rho_{\rm i}=v_{\rm ti}/\omega_{\rm ci}$ is  the ion gyroradius, $v_{\rm ti}=\sqrt{T_{\rm i}/m_{\rm i}}$ is the ion thermal velocity, $\rho_{\rm s}=c_{\rm s}/\omega_{\rm ci}$ is  the ion sound gyroradius, $c_{\rm s}=\sqrt{T_{\rm e}/ m_{\rm i}}$ is the ion sound speed, and $k$ is the wave number, and $\parallel$ and $\perp$ denote the components parallel and perpendicular to the background magnetic field, respectively. The terms $k_{\perp}^2\rho_{\rm s}^2$ and $k_{\perp}^2\rho_{\rm i}^2$ incorporate the effect of the finite perpendicular wavelength into the KAW dispersion relation.

For a typical low temperature laboratory plasma, $T_{\rm i}$ is small and $T_{\rm e} \gg T_{\rm i}$, resulting in $k_{\perp}^2 \rho_{\rm i}^2 \ll 1$ and  $k_{\perp}^2 \rho_{\rm s}^2 \gg k_{\perp}^2 \rho_{\rm i}^2 $.  Under such conditions, Equation (\ref{KAW_dispersion}) reduces to \citep{gekelman1997laboratory, gekelman2011many}
\begin{equation}
    \frac{\omega}{k_{\parallel}}=v_{\rm A}\sqrt{1-\bar{\omega}^{2}+k_{\perp}^2\rho_{\rm s}^{2}}.
    \label{KAW_dispersion2}
\end{equation}
\noindent
The term $1-\bar{\omega}^{2}$ represents the finite frequency correction. It causes the parallel phase velocity, $v_{\rm ph, \parallel} = \omega/k_{\parallel}$, of a shear Alfv{\'e}n wave to decrease as $\omega$ approaches $\omega_{\rm ci}$. When $\bar{\omega}^2 \ll 1$ and $k_{\perp}^{2}\rho_{\rm s}^{2}\ll 1$, the KAW dispersion relation given by Equation (\ref{KAW_dispersion2}) reduces to the ideal MHD shear Alfv{\'e}n wave dispersion relation.

For $\bar{\beta} \ll 1$, the inertia of the electrons becomes important and the wave is called an inertial Alfv{\'e}n wave (IAW). The dispersion relation of an IAW is given by \citep{stasiewicz2000small} 
\begin{equation}
	\frac{\omega}{k_{\parallel}}=v_{\rm A}\frac{\sqrt{\left(1-\bar{\omega}^{2}\right) \left(1+k_{\perp}^2 \rho_{\rm i}^2\right)}}{\sqrt{1+k_{\perp}^2\delta^{2}}} .\label{IAW_dispersion}
\end{equation}
\noindent
Here $\delta={c}/\omega_{\rm pe}$ is the collisionless electron skin depth, $c$ is the speed of light, and $\omega_{\rm pe}=\sqrt{ne^2/m_{\rm e}\epsilon_{0}}$ is the electron plasma frequency, where $e$ is the fundamental unit of electrical charge and $\epsilon_{0}$ is the permitivity of free space. The IAW dispersion relation reduces to ideal MHD shear Alfv{\'e}n wave dispersion relation when $\bar{\omega}^2$, $k_{\perp}^2 \rho_{\rm i}^2 $,  and $k_{\perp}^{2}\delta^2 $ are all $\ll 1$.

KAWs and IAWs propagate both parallel and perpendicular to the ambient magnetic field with finite parallel and perpendicular group velocities  given by $v_{\rm g, {\parallel}}=\partial \omega / \partial k_{\parallel}$ and $v_{\rm g, {\perp}}=\partial \omega / \partial k_{\perp}$, respectively. Typically, $v_{\rm g, \|} \gg v_{\rm g, \perp}$. As a result, energy is transported by these waves predominantly along the magnetic field lines. We also remind the reader that phase velocity refers to velocity of the crest or trough of the wave, while group velocity is the propagation velocity of the total wave envelope.

\subsection{Alfv{\'e}n waves in coronal holes and in LAPD}

Shear Alfv{\'e}n waves are excited by sloshing of the plasma in the photosphere \citep{narain1996chromospheric, priest_2014}. They propagate upward through coronal holes along the ambient magnetic field lines. These waves are in the $\bar{\beta}\gg 1$ regime at low heights and hence, referred to as KAWs. We match this in LAPD in the region where we excite the waves by setting the parameters such as $n$, $T_{\rm e}$ and $B_{0}$ to satisfy the condition $\bar{\beta}\gg 1$ (See Table \ref{table:lab_corona}).  

In coronal holes, most of the wave energy occurs at  $\omega \ll \omega_{\rm ci}$. For example, \cite{morton2015investigating} reported frequencies $f=\omega/2\pi$ of between $0.2 - 16 \;\rm mHz$. The  ambient magnetic field in a coronal hole is $\sim 0.7\;\rm G$ at a height of $0.15\;\rm R_{\odot}$ \citep{morton2015investigating}. At this height, $\bar{\omega}$ ranges from \mbox{ $\approx 1.9\times 10^{-7} - 1.5 \times 10^{-5}$}.   

To the best of our knowledge there are no measurements of $k_{\perp}$ for KAWs in coronal holes. But for the nearly ideal MHD conditions commonly used to model coronal holes, it is typically assumed that both  $k_{\perp}^2 \rho_{\rm i}^2$ and $k_{\perp}^{2}\rho_{\rm s}^2 $ are $\ll 1$. As a consequence of $\bar{\omega}^2$, $k_{\perp}^2 \rho_{\rm i}^2$, and $k_{\perp}^2\rho_{\rm s}^2$ all being $ \ll 1$, shear Alfv{\'e}n waves in  coronal holes are treated as dispersionless, i.e., their frequency varies as $\omega = v_{\rm A}k_{\parallel}$. 

We have designed our shear Alfv{\'e}n wave experiments in LAPD so that $k_{\parallel}$ varies almost linearly with $\omega$ giving $\left( \omega/k_{\parallel}\right) /v_{\rm A} \approx 1$. We also excite shear Alfv{\'e}n waves with dominant perpendicular wavelengths much greater than $\rho_{\rm i}$ and $\rho_{\rm s}$. This ensures that both $k_{\perp}^{2}\rho_{\rm i}^2$ and $k_{\perp}^{2}\rho_{\rm s}^2$ are $ \ll 1$. Additionally, we limit the range of $\bar{\omega}$ from $0.3$ to $0.5$ in order to keep the finite frequency correction as small as possible.  

\begin{deluxetable}{ccCrlc}
	\tablecaption{Dimensionless parameters for coronal holes and LAPD}
	\tablecolumns{6}
	\tablenum{1}
	\tablewidth{0pt}
	\tablehead{
		\colhead{Parameter} &
		\colhead{Coronal hole} &
		\colhead{LAPD}   
	}
	\startdata
	$\bar{\beta}$ & $3 - 18 $  & 1 - 16   \\
	$\bar{\omega}$ & $ \lesssim 1.5 \times 10^{-5}$ & 0.3 - 0.5 \\
	$k_{\perp}^{2} \rho_{\rm i}^{2}$ & $\ll 1^{a}$ & \ll 1   \\
	$k_{\perp}^{2} \rho_{\rm s}^{2}$ & $\ll 1^{a}$ & \ll 1   \\
	$\left(\omega / k_{\parallel}\right) /v_{\rm A}$  & 1 &  \approx 1   \\
	$\beta_{\rm e}$ & $ 1.5 - 9.6 \times 10^{-3}$ & 0.1 - 2.1 \times 10^{-3}    \\
	${\lambda}/{L_{\rm A}}$ & $\gtrsim 4.5 $  & \approx 0.27 - 6.3   \\
	$L_{\rm A}/\lambda_{\rm mfp, e}$ & $\sim 13 $  &  8- 20   \\ 
	$b/B_{0}$ & $\lesssim 0.02$  &  \lesssim 8 \times 10^{-5}   \\
	\enddata
	\tablenotetext{a}{Based on the assumtion that shear Alfv{\'e}n waves in coronal holes satisfy nearly ideal MHD conditions.}
	\label{table:lab_corona}
\end{deluxetable}

In coronal holes, $T_{\rm e} \sim T_{\rm i}$, whereas in LAPD, $T_{\rm e} > T_{\rm i}$. This minor difference does not effect the wave dispersion either in coronal holes or LAPD as the term containing the effect of finite ion temperature in KAW dispersion, $k_{\perp}^2 \rho_{\rm i}^2$,  is negligibly small in both cases.

In coronal holes, magnetic pressure dominates over thermal pressure.  This is represented by the dimensionless parameter $\beta_{\rm e} = 2\mu_0 n T_{\rm e} /B_{0}^2$, where $\mu_0$ is the permeability of free space.  The value of $\beta_{\rm e}$ varies between $\approx 9.6\times 10^{-3}$ and $1.5\times 10^{-3}$ from the surface of the Sun to a height of $0.5\;\rm R_{\odot}$, respectively. To match this in LAPD, we adjusted $B_{0}$, $n$, and $T_{\rm e}$ to produce a  value of $\beta_{\rm e}$ ranging from $\approx 2.1 \times 10^{-3}$ where the waves are excited to $\approx 0.1 \times 10^{-3}$ after the $v_{\rm A}$ gradient. 

\begin{figure}
	\centering
	\includegraphics[scale = 0.4]{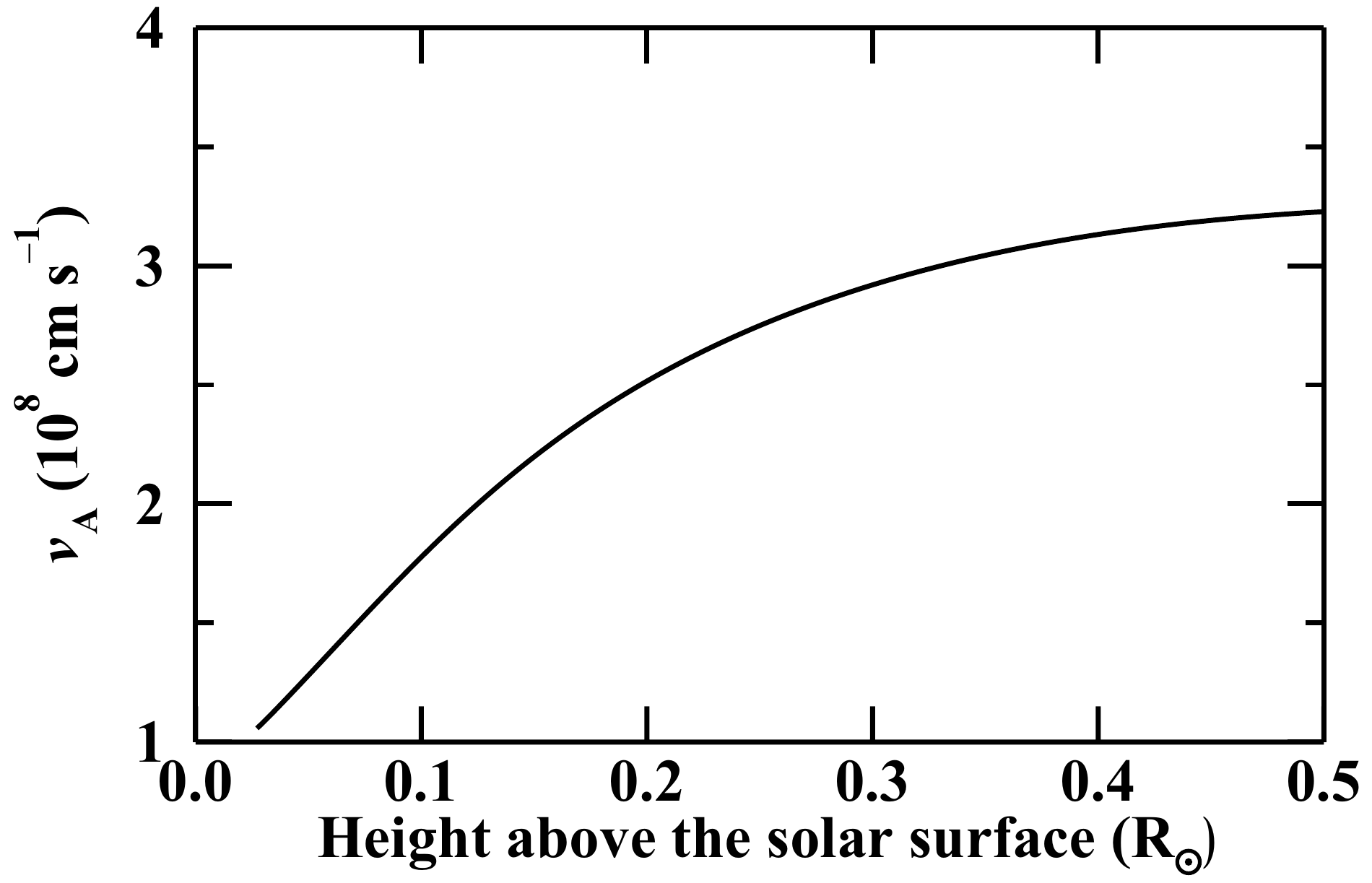}
	\caption{Variation of the Alfv{\'e}n speed, $v_{\rm A}$, in a coronal hole vs.\ height above the solar surface.   
		\label{Alfven_speed_var}}
\end{figure}

In coronal holes, the plasma density and magnetic field are highly non-uniform at low heights. This results in a strong spatial inhomogeneity in $v_{\rm A}$. The predicted spatial variation is shown in Figure \ref{Alfven_speed_var}. The density and magnetic field used here to calculate $v_{\rm A}$ are from the approximate fits given by \citet{cranmer2005generation}. 

For a shear Alfv{\'e}n wave propagating through a longitudinal gradient in $v_{\rm A}$, inhomogeneity-driven effects are predicted to be strong if $v_{\rm A}$ changes substantially over a single wavelength \citep{campos1988oblique,musielak1992klein}. Here, the inhomogeneity parameter can be written as $\lambda/L_{\rm A}$, where $\lambda$ is the wavelength of the shear Alfv{\'e}n wave and $L_{\rm A}$ is the minimum scale length of $v_{\rm A}$ in the gradient. This scale length is defined as  $v_{\rm A}/v^{\prime}_{\rm A}$, where $v^{\prime}_{\rm A}$ is the first spatial derivative of $v_{\rm A}$. For $\lambda / L_{\rm A} \ll 1$, the plasma medium is considered to be homogeneous and for $\lambda / L_{\rm A} \gtrsim 1$, it is considered to be inhomogeneous.

In coronal holes, $L_{\rm A}$ is $\approx \rm 0.1\; R_{\odot}$ (see Figure \ref{Alfven_speed_var}). Alfv{\'e}nic waves have a broad wavelength  spectrum with substantial power in the region of $\lambda \rm \gtrsim 0.45 \;R_{\odot}$  \citep{morton2015investigating}. As a result, the inhomogeneity parameter in coronal holes is predicted to be $\lambda/L_{\rm A} \rm \gtrsim  4.5 $. 

Using LAPD, we have varied $\lambda / L_{\rm A}$ from $\approx 0.27-6.3$. The wavelength of the shear Alfv{\'e}n wave was increased by reducing the frequency of the excited wave. $L_{\rm A}$ was controlled by varying the magnetic field gradient in LAPD. We also note that $\lambda/L_{\rm A} \approx 0.23$ corresponds to wave data acquired in the uniform magnetic field case. In a uniform magnetic field and uniform plasma, $L_{\rm A}$ should be infinite and $\lambda/L_{\rm A} $ should be zero. However, in LAPD, there is a weak variation in density along the axis of the machine that gives rise to an even weaker gradient in $v_{\rm A}$. This effect produces a large but finite value of $L_{\rm A}$ that sets the lower limit for the achievable values of $\lambda/L_{\rm A}$. However, this weak background density variation along LAPD is negligible compared to that due to the magnetic fields applied to generate the $v_{\rm A}$ gradient, as we show below.  Lower values of $L_{\rm A}$ are achieved by increasing the slope of this applied gradient.

In coronal holes, a consequence of the spatial variation of $v_{\rm A}$ is that $\bar{\beta}$ also varies with height. The value varies from $\approx$ 18 at the base of a coronal hole to $\approx$ 3 at a height of $\rm 0.5\;R_{\odot}$. Thus, shear Alfv{\'e}n waves are kinetic at low heights; but the waves are expected to exhibit properties between kinetic and inertial with increasing heights. We have mimicked coronal hole conditions in LAPD by exciting the shear Alfv{\'e}n wave in a region with $\bar{\beta}=16$. The value of $\bar{\beta}$ then approaches 1 as the wave propagates through the gradient in $v_{\rm A}$. 

Shear Alfv{\'e}n waves are also known to damp due to Coulomb collisions \citep{cramer2011physics}. This  damping is predicted to affect shear Alfv{\'e}n waves at low heights in coronal holes \citep{cranmer2002coronal}. The effect of electron-ion collisions on the wave damping in the gradient can be estimated from the ratio of the mean free path of the electrons, $\lambda_{\rm mfp, e}$, to the scale length of the gradient, $L_{\rm A}$. This ratio gives a measure of the number of electron mean free paths within the $v_{\rm A}$ gradient.  The value of $\lambda_{\rm mfp, e}$ was calculated  using
\begin{equation}
    \lambda_{\rm mfp, e}=v_{\rm te} \tau_{ei}=1.46\times10^{11} \frac{T_{e}^2}{nZ_{\rm ch} \ln \Lambda}, \label{Eq:mfp,e}
\end{equation} 
\noindent
where $\tau_{ei}$ is the electron-ion collision time \citep{Braginskii1965}, $Z_{\rm ch}$ is the charge state of the ion, and $\ln \Lambda$ is the Coulomb logarithm \citep{huba2016nrl}. Here, $\lambda_{\rm mfp, e}$ is in m, for $n$ in $\rm{cm^{\rm -3}}$ and $T_{\rm e}$ in eV. For coronal hole conditions of $n\sim 10^{7}\; \rm cm^{-3}$ and $T_{\rm e}\sim 86 \;\rm eV$ ($10^6\;\rm K$) at 0.2 $\rm {R_{\odot}}$, we find $\lambda_{\rm mfp, e}$ is $\sim 5.3\times 10^6 \; \rm{m} = 0.008\;\rm{R_{\odot}} $. Therefore, in coronal holes, $L_{\rm A}/\lambda_{\rm mfp, e} \sim 13$. In LAPD, we have set the value of $n$, $T_{\rm e}$ and $L_{\rm A}$ such that $L_{\rm A}/\lambda_{\rm mfp, e}$ varied from 8 to 20.

Coronal holes extend from the surface of the Sun to interplanetary space, but LAPD is of finite length. However, the magnetic field profile in LAPD is tailored to avoid finite boundary effects. A magnetic beach \citep{stix1992waves} is located between the region where we have performed the experiments and the mechanical boundary of LAPD. Shear Alfv{\'e}n waves are known to damp very strongly in a magnetic beach due to ion cyclotron resonance, thereby preventing the waves from reaching the mechanical boundary of LAPD. Thus, from the perspective of the wave, LAPD looks infinite.

Lastly, Alfv{\'e}nic waves in coronal holes have a range of amplitudes depending on $\omega$ and $k_{\parallel}$ \citep{morton2015investigating}. Waves having normalized amplitude as high as $b/B_{0} \sim 0.02$ were reported by \cite{mcintosh2011alfvenic}. In LAPD, our experiments were restricted to $b/B_{0} \lesssim 8\times 10^{-5}$ .  This low amplitude regime enabled us to avoid known nonlinear effects associated with large amplitude shear Alfv{\'e}n waves. In the  future, we hope to carry out similar experiments in the large amplitude regime.

\section{Experimental overview}\label{EXp}

\subsection{Experimental set up}

\begin{figure*}
	\includegraphics[scale = 0.45]{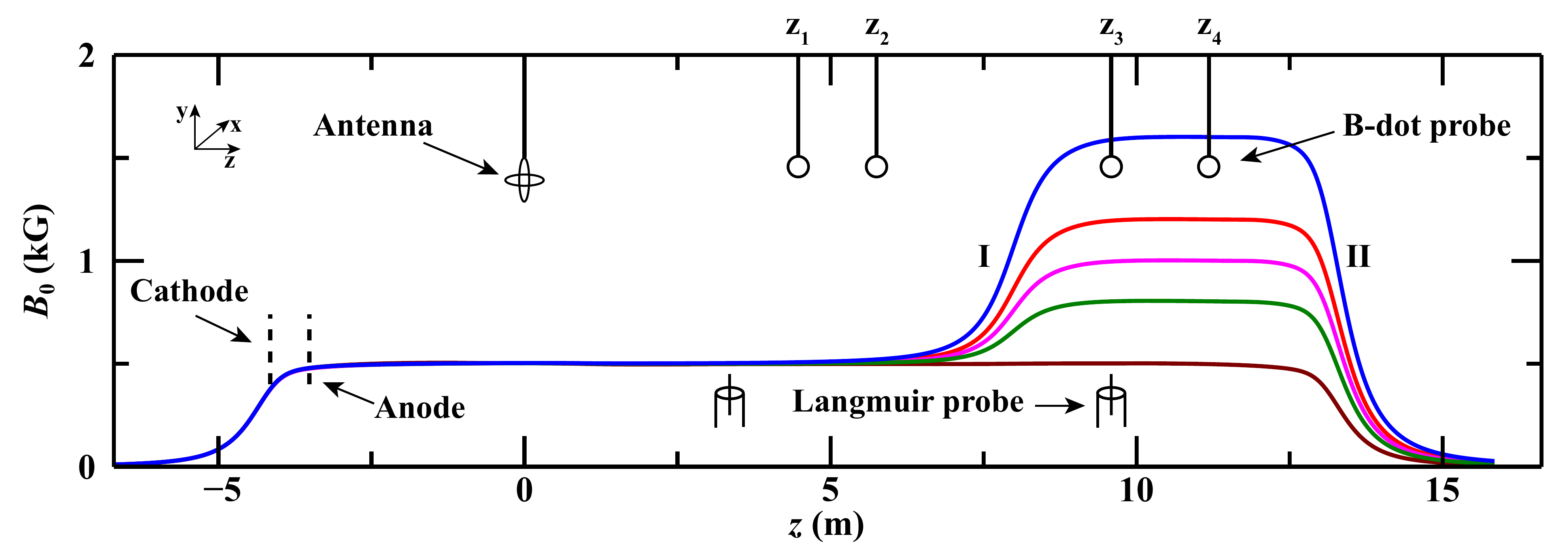}
	\caption{Schematic of the experimental arrangement. The various axial magnetic field profiles used are represented by the different colors. The magnetic field on the low field side, $B_{\rm lo}$ was set to 500 G for all cases. And the magnetic field on the high field side,  $B_{\rm hi}$ was set to 500~G (maroon), 800~G (green), 1000 G (magenta), 1200~G (red), and 1600~G (blue).  The orthogonal ring antenna used to excite shear Alfv{\'e}n waves is centered on the cylindrical axis of LAPD at $x=y=z = 0$. The  vertical ring of the antenna lies in the $yz$ plane, while the horizontal ring lies in the $xz$ plane. The first and second gradient in $v_{\rm A}$ encountered by the excited Alfv{\'e}n waves are labeled as I and II, respectively. Also shown are the diagnostics that were used to measure the plasma parameters and the wave magnetic field. See the text for additional details.      
		\label{exp_set_up}}
\end{figure*}

LAPD houses a 19 m long magnetized plasma column in a cylindrical vacuum chamber of length 24.4 m \citep{gekelman2016upgraded}. The plasma was produced by applying a voltage between a 60 cm diameter hot barium oxide cathode \citep{leneman2006plasma} and a mesh anode located 0.5~m away, as shown in Figure \ref{exp_set_up}. Each plasma discharge was pulsed, and was operated at a 1~Hz repetition rate. The duration of each discharge or shot was $\approx 10~\rm ms$.  The experiments were performed in a helium plasma. The neutral  helium pressure was held constant at $ \sim \rm 10^{-4}\;torr$. 

Ten sets of electromagnets are arranged coaxially with the vacuum chamber in order to produce the axial magnetic field. The axial magnetic field points in the $- \hat{z}$ direction for the coordinate system adopted in this paper.  We controlled $\lambda/L_{\rm A}$, in part, by creating a gradient in the axial magnetic field.  On the low field side of the machine, we set the magnetic field, $B_{\rm lo}$, to 500 G.  $L_{\rm A}$ was varied by setting the field strength on the high field side, $B_{\rm hi}$, to one of five different values: 500, 800, 1000, 1200, and 1600~G. These magnetic field profiles are shown in Figure \ref{exp_set_up}. 

Shear Alfv{\'e}n waves were excited using an orthogonal ring antenna located on the axis of LAPD at $x=y=z=0$ \citep{gigliotti2009generation, karavaev2011generation}. 
The diameter of the ring is 9~cm. The dominant perpendicular wavelength, $\lambda_{\perp}$, excited by the orthogonal antenna was typically $\sim 28\;\rm cm$. This value of $\lambda_{\perp}$ was determined from the wave data using a Fourier-Bessel analysis \citep{churchill1987introduction} as illustrated by \cite{vincena1999}. In our experiments the dominant $\lambda_{\perp}$ was large enough to ensure that $k_{\perp}^{2}\rho_{\rm i}^{2}$, $k_{\perp}^{2}\rho_{\rm s}^{2}$, and  $k_{\perp}^{2}\delta^{2}$ are all $\ll 1$.  
 
The Alfv{\'e}n wave magnetic fields were measured using triaxial B-dot probes. Each probe consist of three oppositely wound orthogonally oriented coils. The signals from each pair of coils were amplified using a differential amplifier to avoid electrostatic pick up. The amplified signal was averaged over 14 shots and digitized using a 16-bit data acquisition system. This allows us to detect wave magnetic fields as small as 0.5~mG. The probes are mounted on computer-controlled $xy$ translators that enabled us to map out the wave magnetic field along a cross section of LAPD. The B-dot probes used for most of the measurements reported here were located at axial distances of $\rm z_{1} = 4.47\;\rm  m$, $\rm z_{2} = 5.75 \;\rm m$, $\rm z_{3}=9.59\;\rm m$, and $\rm z_{4}=11.18\rm \; m$.

\subsection{Equilibrium plasma parameters}

\begin{figure}
	\centering
	\includegraphics[scale = 0.26]{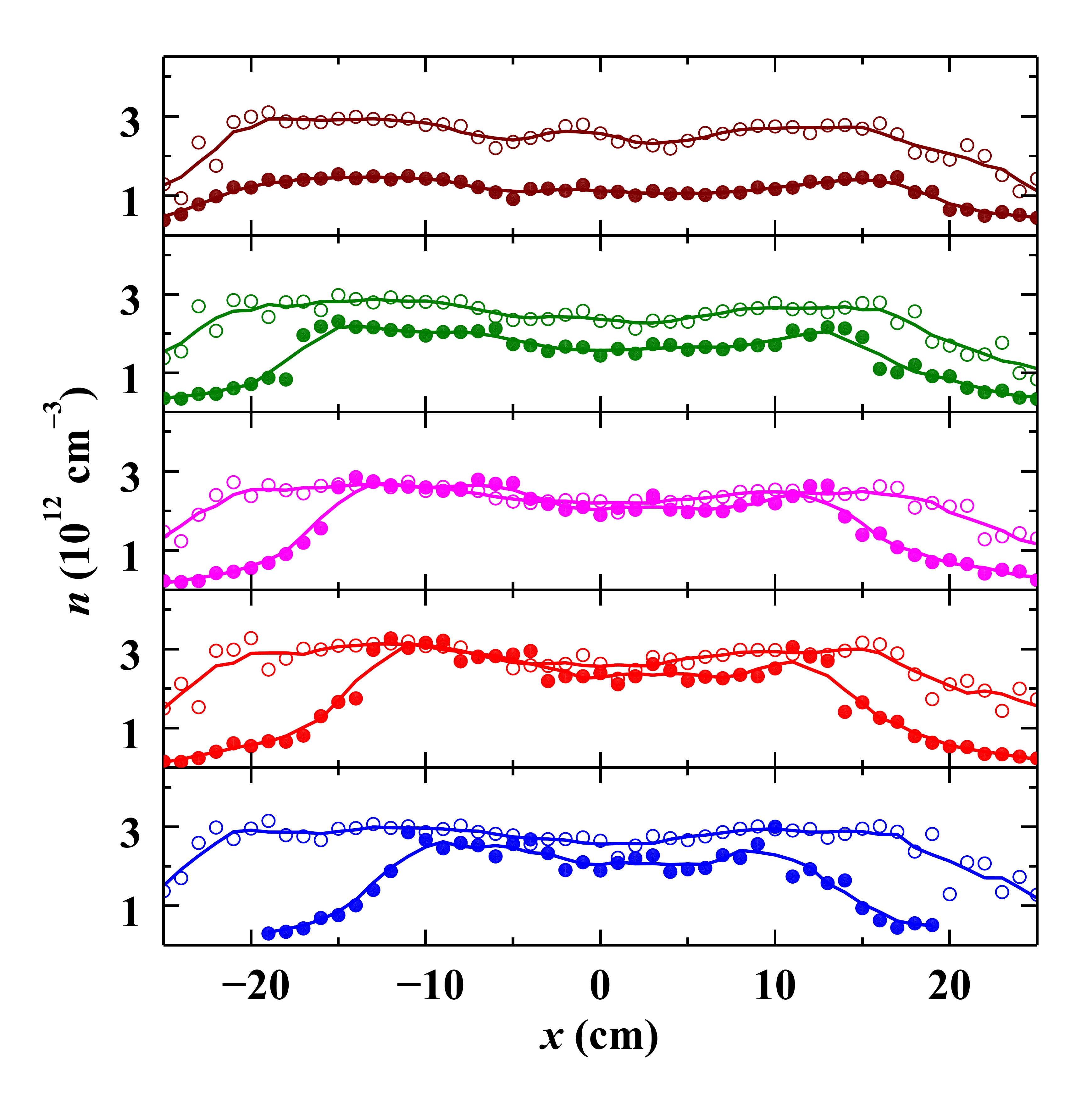}
	\caption{Variation of the plasma density for $y=0$ along the $x$ axis for the applied axial magnetic field configurations. The different colors represent the magnetic field profiles shown in Figure \ref{exp_set_up}. The open and filled circles show the data acquired by the Langmuir probe before the magnetic field gradient at $z$ = 3.50~m and after at 9.59~m, respectively.  
		\label{ni}}
\end{figure}

Plasma parameters, such as $T_{\rm e}$ and $n$, were measured using a Langmuir probe. $T_{\rm e}$ was determined from the slope of the linear region of the $\ln \left(I_{\rm pr,e}\right)$ vs.\ $V_{\rm pr}$ curve, where $I_{\rm pr,e}$ is the electron current collected by the probe and $V_{\rm pr}$ is the probe potential. The ion saturation current collected by the probe was used to determine $n$ after calibrating the probe with a heterodyne microwave interferometer. 

The variation of $n$ along the $x$ axis in the $y = 0$ plane was measured at $z = 3.50\;\rm m$ before the gradient and at $z=9.59\;\rm m$ after the gradient. Figure \ref{ni} shows our measurements for different magnetic field profiles. The uncertainty in each measurement is typically equal to the size of the symbol. Here and throughout the paper, all uncertainties are given at an estimated one-sigma statistical confidence level. 

\begin{deluxetable*}{cccccccc}
    \centering
	\tablecaption{Equilibrium plasma parameters before and after the magnetic field gradient.\label{table:density_temp}}
	\tablecolumns{6}
	\tablenum{2}
	\tablewidth{6pt}
	\tablehead{
		\multicolumn{2}{c}{Magnetic field strength}   &
		\multicolumn{3}{c}{Before gradient}   &
		\multicolumn{3}{c}{After gradient}   \\
		\cmidrule(rl){1-2} \cmidrule(lr){3-5} \cmidrule(lr){6-8}
		 \multicolumn{1}{c}{$\quad \quad B_{\rm lo}$} & \multicolumn{1}{c}{$B_{\rm hi}$} &  \multicolumn{1}{c}{$n$}  & \multicolumn{1}{c}{$T_{\rm e}$}  & \multicolumn{1}{c}{$\bar{\beta}$} & \multicolumn{1}{c}{$n$} & \multicolumn{1}{c}{$T_{\rm e}$}  & \multicolumn{1}{c}{$\bar{\beta}$} \\
		  $\quad\quad$(G)&  (G)&   ($10^{12}\; \rm cm^{-3}$) &  (eV) & &  ($10^{12}\; \rm cm^{-3}$) &  (eV) &  
	}
	\startdata
	$\quad\quad$ 500 & 500 & $2.6\pm 0.3$ & $5.0\pm 0.5 $& 15 & $1.2\pm 0.2  $ & $3.4 \pm 0.4$ & 5   \\
	$\quad\quad 500$ & $800 $ & $2.6\pm 0.3 $ & $5.1\pm 0.5 $ & 16 & $1.8 \pm 0.3$ & $ 3.3 \pm 0.5 $ & 3 \\
	$\quad\quad 500$ & $1000 $ & $2.4\pm 0.2  $ & $4.9 \pm 0.5 $ & 14  & $2.3 \pm 0.3$ & $ 3.3 \pm 0.5 $ & 2 \\
	$\quad\quad500$ & $1200$  & $2.8\pm 0.3  $ & $4.9 \pm 0.5$ & 16 & $2.6 \pm 0.4  $ & $ 3.2 \pm 0.5 $ & 2 \\
	$\quad\quad500$ & $ 1600 $ & $2.8 \pm 0.3 $  & $4.9 \pm 0.5 $ & 16 & $2.3 \pm 0.3$ & $ 3.1 \pm 0.5 $ & 1\\ 
	\enddata
\end{deluxetable*}

For our analysis of the wave data, we used the spatial average of $n$ over the region sampled by the wave. In the low field region, at $z=3.50\;\rm m$, this region spans $ - 20 \; \rm{cm} \leq \mathit{x} \leq 20 \;\rm{cm}$ (as shown later in Section \ref{Obs}). On the high field region, the cross section of the plasma sampled by the Alfv{\'e}n waves decreases due to  the convergence of magnetic field lines. This decreased sample region is determined using the flux conservation equation, $x_{\rm hi} =\sqrt{{B_{\rm lo} x_{\rm lo}^2}/{B_{\rm hi}}}$, where $B_{\rm lo}$ = 500~G, $x_{\rm lo}$ = 20~cm, and $B_{\rm hi}$ is value of the magnetic field at $z = 9.59 \; \rm m$. The averaged $n$ and $T_{\rm e}$ before and after the gradient for different magnetic field configurations are given in Table \ref{table:density_temp}. The ion temperature in LAPD was typically $\rm \sim 1 \;eV$.

\section{Wave experiment results}\label{Obs}

\subsection{Excitation of shear Alfv{\'e}n waves}\label{SAW_dispersion}

\begin{figure}
	\centering
	\includegraphics[scale = 0.55]{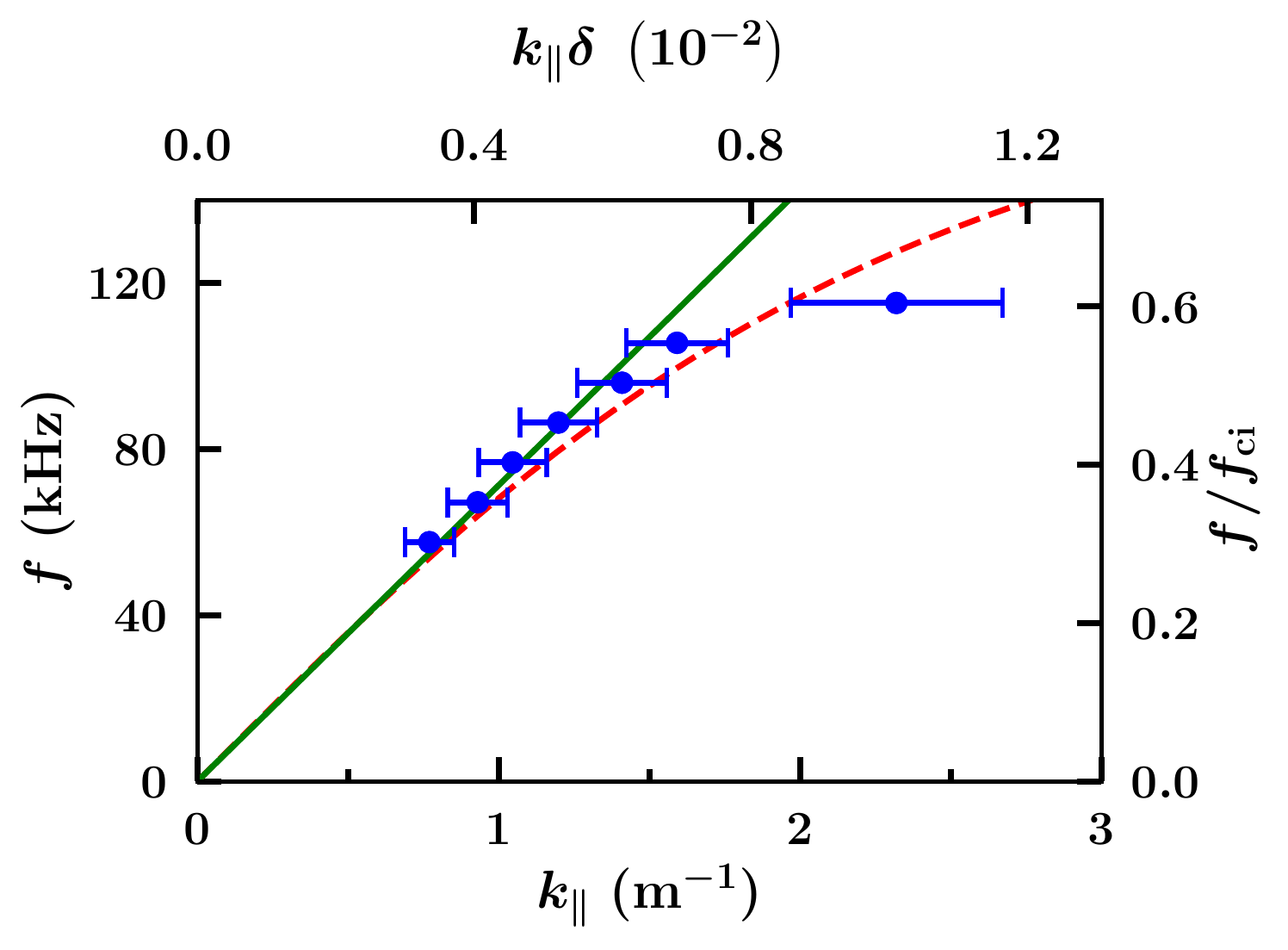}
	\caption{Dispersion relation of the shear Alfv{\'e}n waves in a uniform 500 G axial magnetic field. The blue symbols represent the experimental data. The green solid and red dashed lines plot the dispersion relation for ideal MHD shear Alfv{\'e}n waves and for KAWs, respectively. The $y$ axis on the left gives the wave frequency and on the right shows the wave frequency as a fraction of the ion cyclotron frequency. The lower $x$ axis is the parallel wave number while the upper $x$ axis shows the parallel wave number in terms of the dimensionless quantity $k_{\parallel} \delta$. 
		\label{dispersion_500G}}
\end{figure}

Linearly polarized shear Alfv{\'e}n waves were excited by applying a sinusoidal wave train of ten cycles to the horizontal ring of the antenna. The dispersion relation of the excited shear Alfv{\'e}n wave is shown in Figure \ref{dispersion_500G}. The quantity $k_{\parallel}$ was measured from the phase difference in the wave magnetic field between $\rm z_{1}$ and $\rm z_{2}$.          

The measured value of $k_{\parallel}$ varies nearly linearly with $f$ for $f \leq 0.5f_{\rm ci}$. 
 Following the predictions of ideal MHD, we have fit a straight line to the data for $f \leq 0.5f_{\rm ci}$. The value of $v_{\rm A}$ determined from the slope of the fitted  line is found to be within $\rm 14\%$ of that calculated using $n$ measured with the Langmuir probe. This minor disagreement we attribute to the cumulative uncertainties in the $k_{\parallel}$  and Langmuir probe measurements. 

The theoretical KAW dispersion relation given by Equation (\ref{KAW_dispersion}) is presented by the dashed curve in Figure \ref{dispersion_500G}. The value of $v_{\rm A}$ obtained from the fitted straight line is used to calculate this dashed curve. The measured variation of $f$ vs.\ $k_{\parallel}$  is found to be in good agreement with Equation (\ref{KAW_dispersion}), confirming that the waves excited are indeed KAWs. 

\subsection{Propagation through a longitudinal gradient in the Alfv{\'e}n speed} \label{SAW_propagation}

\subsubsection{Wave properties before and after the gradient}\label{subsec:bf_af}

The value of $\overline{\beta}$ is $\gg 1$ in the low field region, where the shear Alfv\'en waves are excited, and decreases to $1$ in the high field region (see Table \ref{table:density_temp}). As a result, the excited shear Alfv{\'e}n waves do not strictly match to the definition of KAWs at all points in space. Hence, we use the more general term shear Alfv{\'e}n waves to refer to the waves excited by the antenna.

\begin{figure*}
	\centering
	\includegraphics[scale = 0.45]{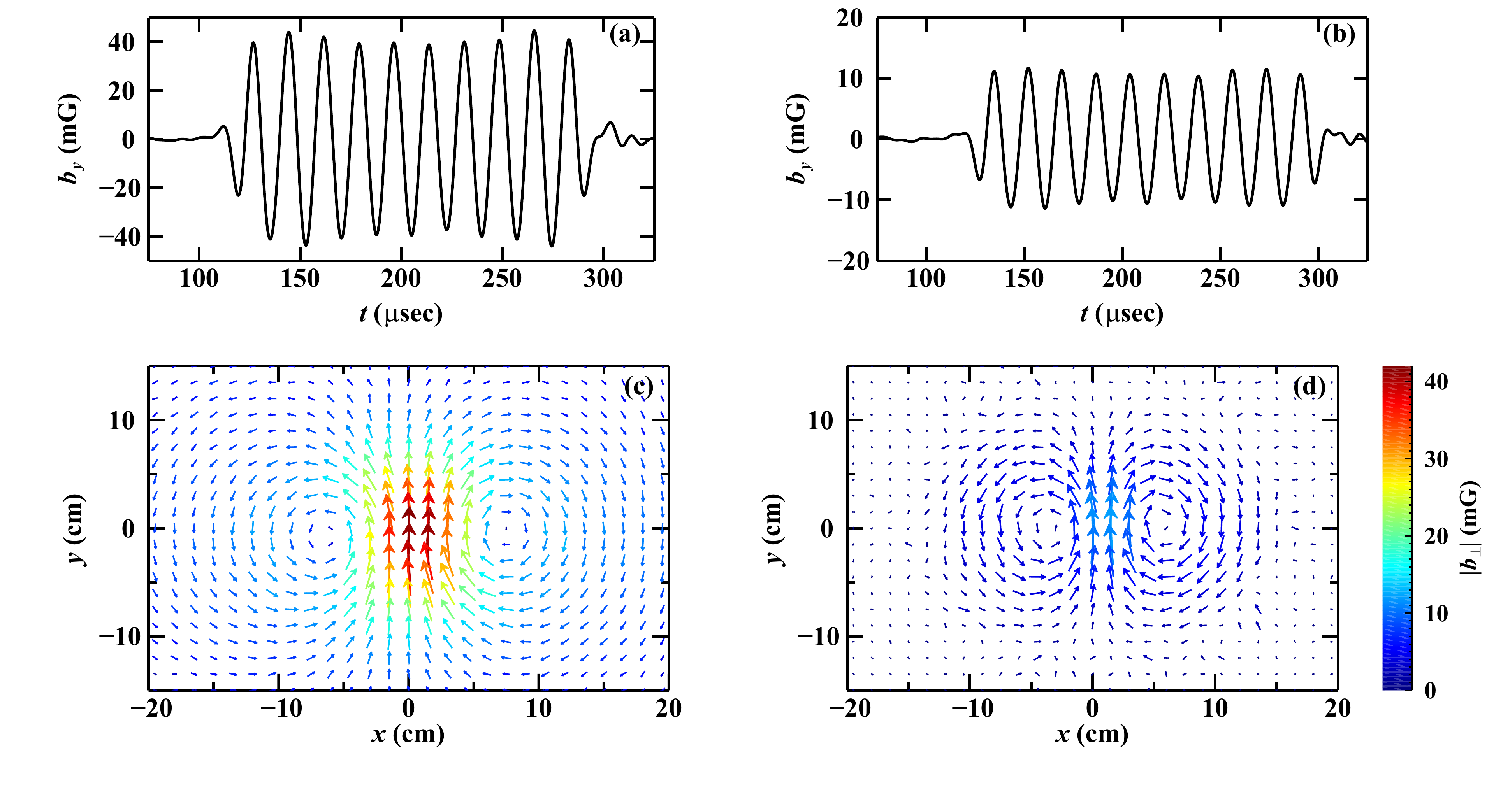}
	\caption{The time variation of the $y$ component of the shear Alfv{\'e}n wave magnetic field on the axis of LAPD at (a) $\rm z_{2}$ and (b) $\rm z_{4}$, respectively. Also shown are the spatial variation of the wave magnetic vector fields in the $xy$ cross section of LAPD at (c) $\rm z_{2}$ and (d) $\rm z_{4}$ at times of $t$ = $161.8$, and $169.2 \; \rm \mu s$, respectively, corresponding to the third peak of the applied wave train. The direction of the arrows represent that of the wave magnetic field and the colors give the magnitude of the field using the color bar shown. The arrow lengths are normalized by the maximum value of the magnetic field in each panel. For these measurements $B_{\rm lo}$ was held at 500~G and $B_{\rm hi}$ was 1600 G. See text for additional details.         
		\label{vector_2D}}
\end{figure*}

The measured $y$ component of the wave magnetic field is shown in Figure 5 before (a) and after (b) the $v_{\rm A}$ gradient.  The data are shown for $x=y=0$,  $f=57.6$~kHZ, and $B_{\rm hi}=1600$~G.  Figures \ref{vector_2D} (c) and (d) show the structure of the shear Alfv{\'e}n wave  on each side of the magnetic field gradient, respectively. Two well formed current channels are observed with the separation between the current channels being smaller on the high field side of the gradient, as is expected for shear Alfv{\'e}n waves propagating along converging magnetic field lines. 

In order to confirm that the measured property of the shear Alfv\'en wave is in agreement with the expected theoretical value, we have tried to measure the parallel component of the wave magnetic field, $b_{\parallel}$, on the axis of LAPD before and after the gradient.  According to theory \citep{hollweg1999kinetic}, before the gradient for the dominant $k_{\perp} \rho_{\rm s}$ of $\approx$ 0.21 the predicted value of $b_{\parallel}$ is $\approx$ 0.22~mG, while after the gradient for the dominant $k_{\perp}\rho_{\rm s}$ of $\approx$ 0.08 the predicted value of $b_{\parallel}$ is $\approx$ 0.01~mG.  These values of $b_{\parallel}$ are below our measurement threshold of 0.5~mG. Therefore, the lack of detection of a $b_{\parallel}$ is consistent with the theoretical prediction.

\subsubsection{Determination of wave energy}

The energy of the shear Alfv{\'e}n wave is obtained using the Poynting vector, $S$, crossing a plane perpendicular to the ambient magnetic field. This is given by \citep{ karavaev2011generation}
\begin{equation}
	S = \frac{1}{\mu_{0}}b^2v_{\rm g, \parallel}=\frac{1}{\mu_{0}}b^2v_{\rm ph, \parallel}. \label{Eq:poynting}
\end{equation}
\noindent
$S$ is the energy flux. In Equation (\ref{Eq:poynting}), $v_{\rm g, {\parallel}}$ is considered to be equal to $v_{\rm ph, {\parallel}}$ because the experiments are limited to $\omega \leq 0.5 \omega_{ci}$ , where the shear Alfv{\'e}n wave dispersion relation is nearly linear. Hence, the total wave energy, $\mathcal{E}$, passing through the cross section of LAPD perpendicular to the ambient magnetic field can be expressed as 
\begin{equation}
\mathcal{E} = \int \left( \iint S \;\mathrm{d}x \;\mathrm{d}y\right) \mathrm{d}\mathit{t} =  \frac{v_{\rm{ph,} {\parallel}}}{4\pi} \int \left(\iint b^{2} \;\mathrm{d}x \;\mathrm{d}y \right) \mathrm{d}t. \label{Eq:wave_energy}
\end{equation}
\noindent
The spatial integration is carried out over the cross section of LAPD and the integration in time is carried out over the duration of the wave train, examples of which are shown in Figure \ref{vector_2D}. The wave power, $\mathit{\Gamma}$, is related to the total wave energy by the relation
\begin{equation}
\mathit{\Gamma} =  \frac{\mathcal{E}} {t_{\rm dur}},
\end{equation}
\noindent
 where $t_{\rm dur}$ is the duration of the wave train. The value of $v_{\rm ph, {\parallel}}$ used in Equation (\ref{Eq:wave_energy}) was determined by simultaneously measuring the wave magnetic field using two axially separated B-dot probes. The probes were carefully aligned to ensure that both intersected the same axial magnetic field line. To calculate $v_{\rm ph, {\parallel}}$, the axial distance between the probes was divided by the time lag between the phase of the wave magnetic field. The time lag was determined by a cross-correlation analysis of the time variation of the data acquired by the two probes. The probes located at $\rm z_{1}$ and $\rm z_{2}$ were used to measure $v_{\rm ph, {\parallel}}$ in the low field side, while the probes located at $\rm z_{3}$ and $\rm z_{4}$ were used to measure $v_{\rm ph, {\parallel}}$ in the high field side.  

\subsubsection{Reduction in power of the transmitted Alfv{\'e}n wave}\label{Sec:transmitted_energy_reduction}

The reduction in power of the wave propagating through the gradient was measured using the ratio of the transmitted wave power $\mathit{\Gamma}_3$ at z$_3$ to the incident wave power $\mathit{\Gamma}_{2}$ at $z_{2}$. Since, $\mathit{\Gamma}$ is related to $\mathcal{E}$ by a constant factor, the reduction in wave power is equal to the decrease in wave energy, i.e.,  $\mathit{\Gamma}_{3}/\mathit{\Gamma}_{2}=\mathcal{E}_{3}/\mathcal{E}_{2}$. The dependence of $\mathit{\Gamma}_{3}/\mathit{\Gamma}_{2}$ on $\lambda/L_{\rm A}$ was studied by varying $\lambda/L_{\rm A}$ in two ways. In the first set of experiments, $L_{\rm A}$ was changed while holding $\lambda$ constant. In the second set, $\lambda$ was varied and $L_{\rm A}$ was kept constant.

\begin{figure}
	\centering
	\includegraphics[scale = 0.4]{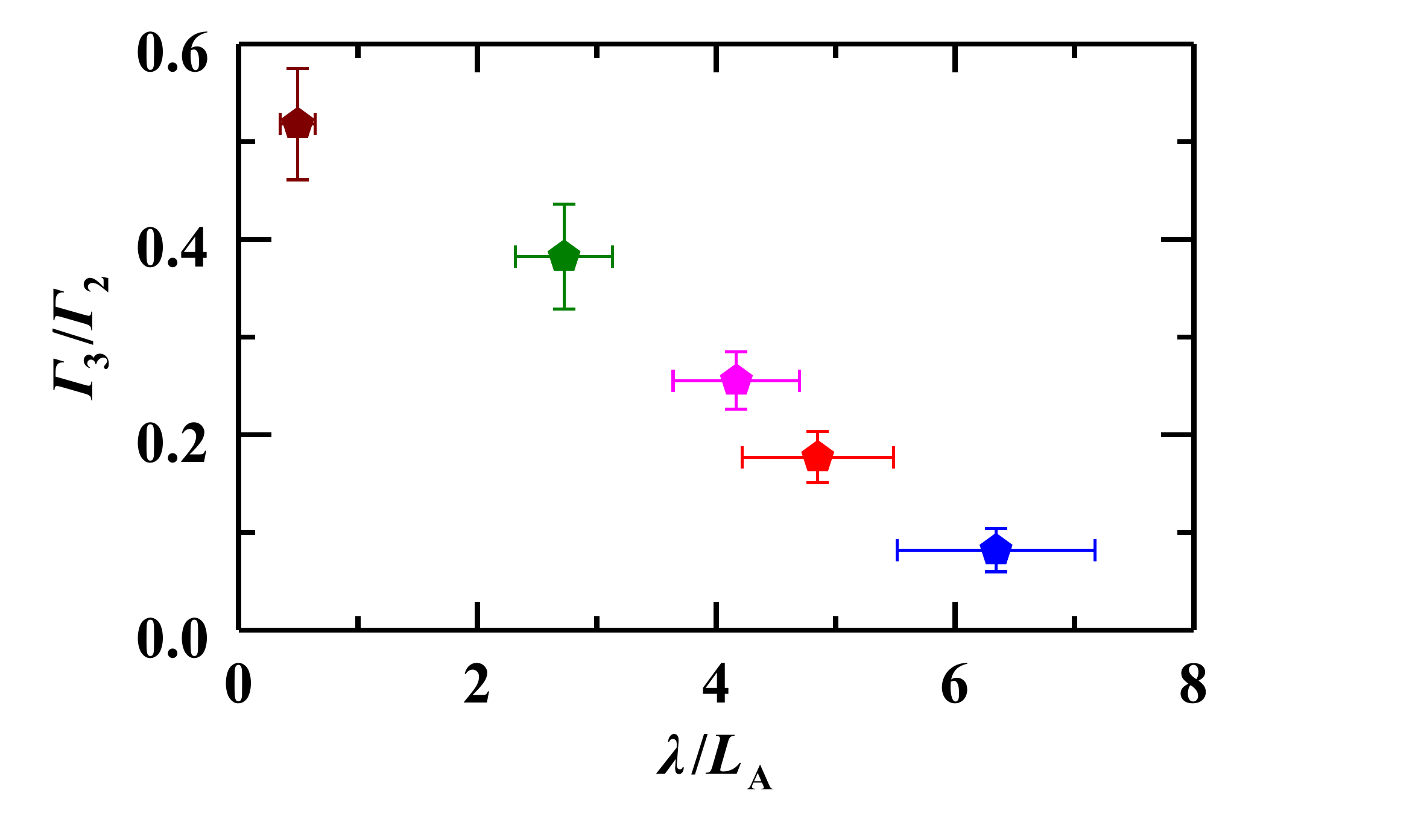}
	\caption{The ratio of the wave power, $\mathit{\Gamma}_{3}/\mathit{\Gamma}_{2}$, passing through the cross section of LAPD at $\rm z_{3}$ and $\rm z_{2}$, respectively, is shown vs.\ $\lambda/L_{\rm A}$. $L_{\rm A}$ was varied by increasing the value of $B_{\rm hi}$. The different colors indicate the different values of $B_{\rm hi}$ as given in Figure \ref{exp_set_up}. The wave frequency was held constant at 57.6 kHz.          
		\label{SL_var}}
\end{figure}

For the first set of experiments, the values of $L_{\rm A}$ were varied by increasing $B_{\rm hi}$ from 500 G to 800, 1000, 1200, and 1600~G. The increase in the value of $B_{\rm hi}$ enhances the steepness of the gradient in $v_{\rm A}$. This, in turn, decreases $L_{\rm A}$.

The variation of $\mathit{\Gamma}_{3}/\mathit{\Gamma}_{2}$ with $\lambda/L_{\rm A}$ while varying $L_{\rm A}$ is shown in Figure \ref{SL_var}. Waves of frequency $f = 57.6$ kHz were excited to keep $\lambda$ fixed. For the nearly homogeneous case of   $\lambda/L_{\rm A}\approx 0.49$, we find $\mathit{\Gamma}_{3}/\mathit{\Gamma}_{2} \; \mathrm{is} \; \approx 0.52$. For a large non-uniformity of $\lambda/L_{\rm A} \approx 6.3$, we find $\mathit{\Gamma}_{3}/\mathit{\Gamma}_{2} \; \mathrm{is} \; \approx 0.08$. These results show that the wave power propagating through the gradient decreases as the steepness of the gradient increases.

\begin{figure}
	\centering
	\includegraphics[scale = 0.39]{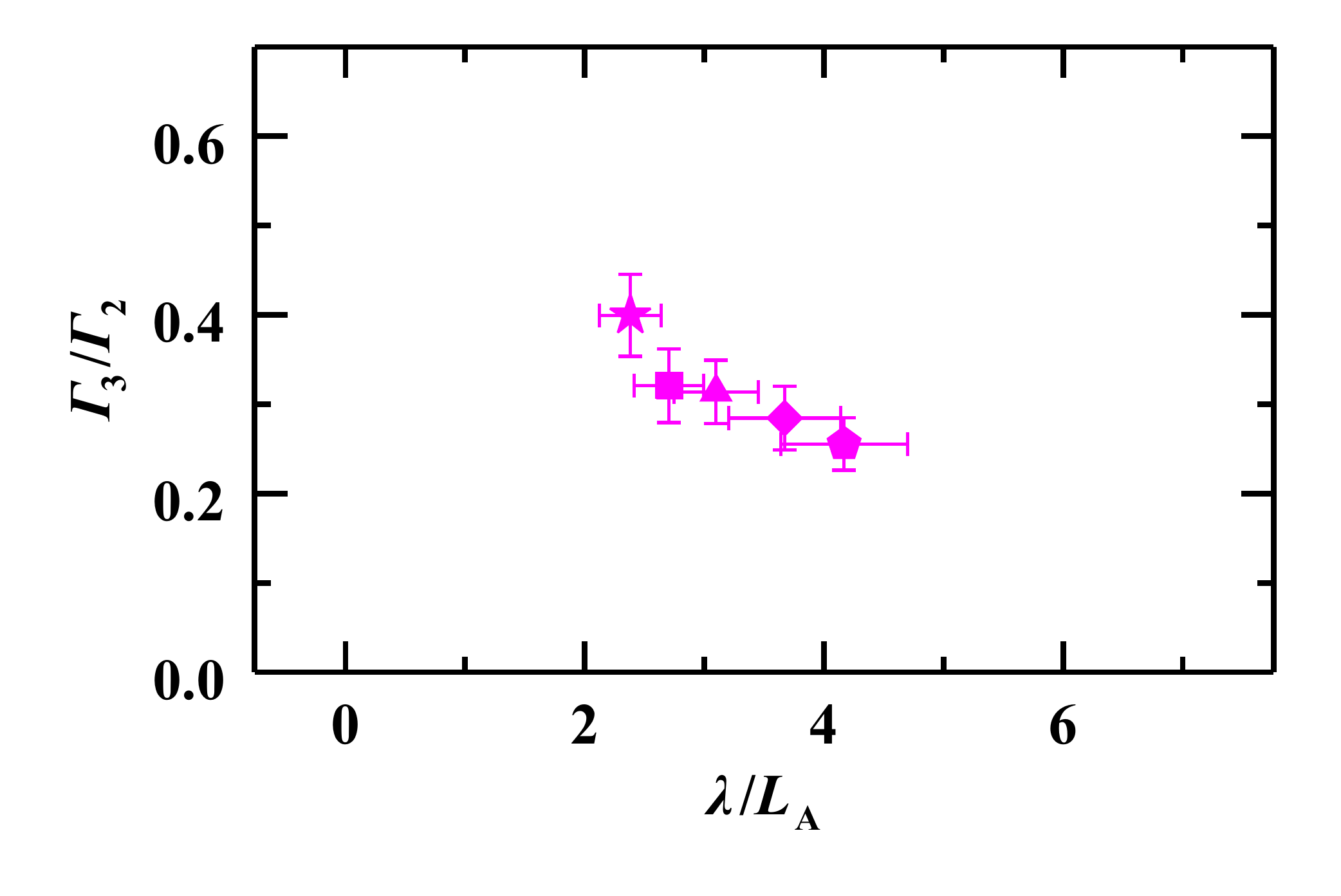}
	\caption{Same as Figure \ref{SL_var}, but here $\lambda/L_{\rm A}$ was varied by changing $\lambda$, while $L_{\rm A}$ was kept fixed with $B_{\rm hi}=1000\rm~G$.  Results are shown for $f$ = 57.6~kHz (pentagon), 67.2~kHz (diamond), 76.8~kHz (triangle), 86.4~kHz (square) and 96.0~kHz (star). 
		\label{F_var}}
\end{figure}

In the second set of measurements, $\lambda$ was varied by changing $f$ from 57.6 kHz to 96 kHz in steps of 9.6 kHz.  We confined $f$ to this range so that the shear Alfv{\'e}n waves followed the linear dispersion relation to a good approximation. $L_{\rm A}$ was kept fixed by setting $B_{\rm hi}$ = 1000 G. Figure \ref{F_var} shows that $\mathit{\Gamma_{\rm 3}/\Gamma_{\rm 2}}$ decreases with increasing $\lambda/L_{\rm A}$ while varying $\lambda$. For example, for $\lambda/L_{\rm A}\approx 2.38$, $\mathit{\Gamma}_{3}/\mathit{\Gamma}_{2} \; \mathrm{is} \; \approx 0.40$, while for $\lambda/L_{\rm A} \approx 4.17$, $\mathit{\Gamma}_{3}/\mathit{\Gamma}_{2} \; \mathrm{is} \; \approx 0.26$. This shows that a wave with a longer wavelength loses more energy than that with a shorter wavelength while propagating through a constant gradient.

\begin{deluxetable}{cccc}[h]
	\tablenum{3}
	\tablecaption{$\mathit{\Gamma}_{3}/\mathit{\Gamma}_{2}$ vs.\ $\lambda/L_{\rm A}$}
	\label{tab:energy_ratio}
	\tablecolumns{4}
	\tablewidth{0pt}
	\tablehead{
		\colhead{$B_{\rm hi}\left(\rm G\right)$} &
		\colhead{$f\left(\rm kHz\right)$} &
		\colhead{$\lambda/L_{\rm A}$} &
		\colhead{$\mathit{\Gamma}_{3}/\mathit{\Gamma}_{2}$}   
	}
	\startdata
	500 & 57.6 & $0.49\pm 0.15$  & $0.52 \pm 0.06$   \\
	500 & 67.2 & $0.41 \pm 0.12$ & $0.56 \pm 0.08$ \\
	500 & 76.8 & $0.36 \pm 0.11$ & $0.54 \pm 0.07$ \\
	500 & 86.4 & $0.32 \pm 0.10$ & $0.42 \pm 0.05$ \\
	500 & 96.0 & $0.27 \pm 0.08$ & $0.40 \pm 0.05$ \\
	800 & 57.6 & $2.73 \pm 0.41$ & $0.38 \pm 0.05$  \\
	800 & 67.2 & $2.27 \pm 0.35$ & $0.39 \pm 0.06$  \\
	800 & 76.8 & $1.88 \pm 0.28$ & $0.42 \pm 0.06$ \\
	800 & 86.4 & $1.69 \pm 0.24$ & $0.41 \pm 0.05$ \\
	800 & 96.0 & $1.54 \pm 0.22$ & $0.43 \pm 0.06$ \\
	1000 & 57.6 & $4.17 \pm 0.53$ & $0.26 \pm 0.03$  \\
	1000 & 67.2 & $3.67 \pm 0.47$ & $0.28 \pm 0.04$  \\
	1000 & 76.8 & $3.10 \pm 0.35$ & $0.31 \pm 0.04$ \\
	1000 & 86.4 & $2.71 \pm 0.29$ & $0.32 \pm 0.04$ \\
	1000 & 96.0 & $2.38 \pm 0.26$ & $0.40 \pm 0.05$ \\	
	1200 & 57.6 & $4.85 \pm 0.64$ & $0.18 \pm 0.03$  \\
	1200 & 67.2 & $4.27 \pm 0.53$ & $0.22 \pm 0.03$  \\
	1200 & 76.8 & $3.80 \pm 0.49$ & $0.23 \pm 0.03$ \\
	1200 & 86.4 & $3.25 \pm 0.35$ & $0.27 \pm 0.04$ \\
	1200 & 96.0 & $2.87 \pm 0.33$ & $0.38 \pm 0.05$ \\
	1600 & 57.6 & $6.34 \pm 0.83$ & $0.08 \pm 0.02$  \\
	1600 & 67.2 & $4.87 \pm 0.61$ & $0.12 \pm 0.03$  \\
	1600 & 76.8 & $4.66 \pm 0.59$ & $0.15 \pm 0.04$ \\
	1600 & 86.4 & $4.24 \pm 0.46$ & $0.17 \pm 0.03$ \\
	1600 & 96.0 & $3.69 \pm 0.43$ & $0.21 \pm 0.04$ \\	
	\enddata
\end{deluxetable}

The decrease in $\mathit{\Gamma_{\rm 3}/\mathit{\Gamma_{\rm 2}}}$ with increasing $\lambda/L_{\rm A}$ presented in Figure \ref{F_var} shows the same quantitative behavior as seen in Figure \ref{SL_var}. This confirms that it is neither $\lambda$ nor $L_{\rm A}$ but rather $\lambda/L_{\rm A}$ that is the independent parameter describing the effect of inhomogeneity on the shear Alfv{\'e}n waves.    

\begin{figure}[b]
	\centering
	\includegraphics[scale = 0.41]{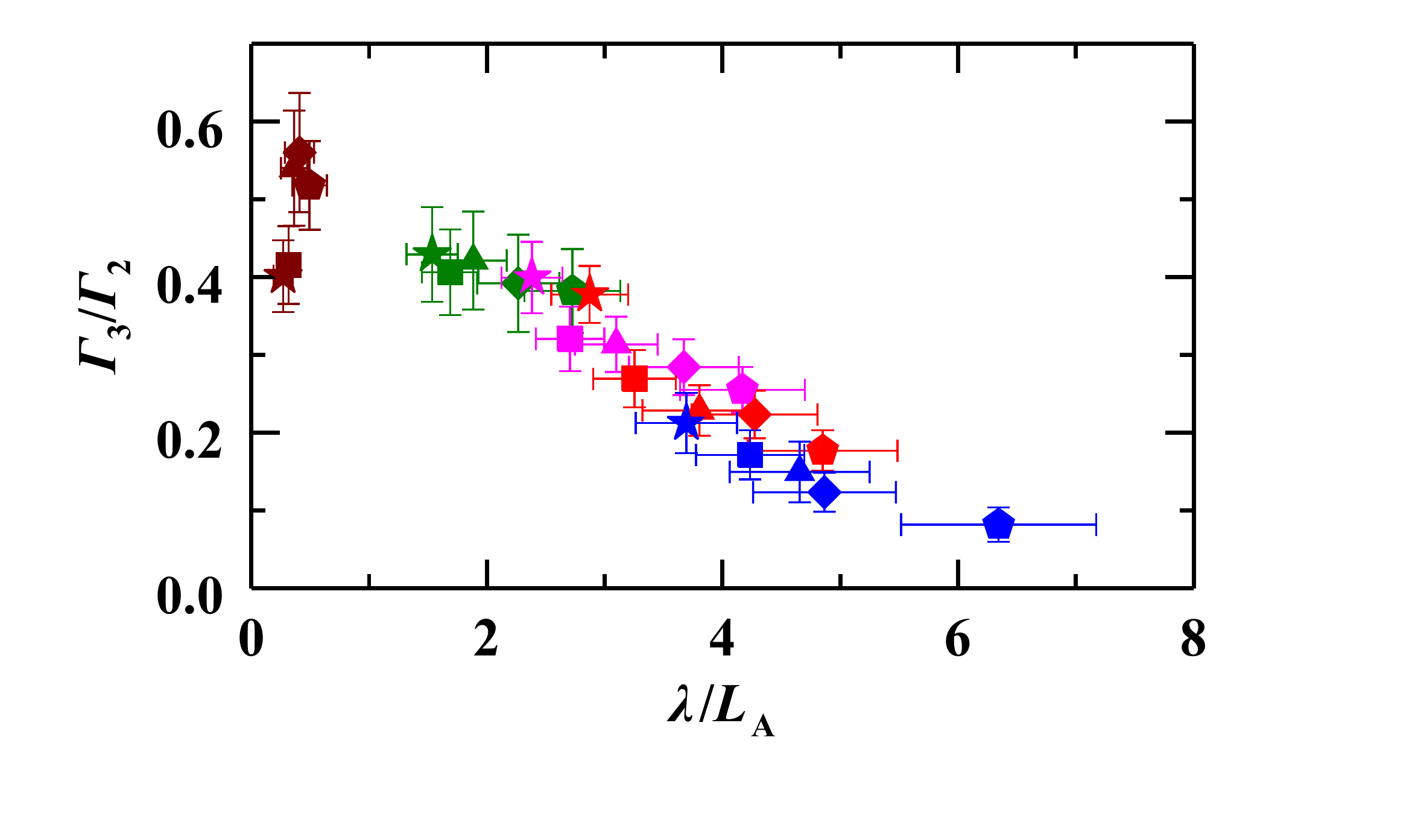}
	\caption{$\mathit{\Gamma_{\rm 3}}/\mathit{\Gamma_{\rm 2}}$ vs.\ $\mathit{\lambda}/L_{\rm A}$ when either $\lambda$ or $L_{\rm A}$ are varied. The colors and symbols are defined in Figures \ref{exp_set_up} and \ref{F_var}, respectively. The plotted data are also given in Table \ref{tab:energy_ratio}. 
		\label{Fig:transmission}}
\end{figure}

The variation of $\mathit{\Gamma_{\rm 3}}/\mathit{\Gamma_{\rm 2}}$ for all measured values of $\lambda/L_{\rm A}$ is shown in Figure \ref{Fig:transmission} and given in Table \ref{tab:energy_ratio}. Here, $\lambda/L_{\rm A}$ was varied by  changing $f$ from 57.6 to 96 kHz for each of the five values of $B_{\rm hi}$ given in Figure \ref{exp_set_up}. The data in Figure \ref{Fig:transmission} include all of the data plotted in Figures \ref{SL_var} and \ref{F_var} as well as the additional values listed in Table \ref{tab:energy_ratio} but not included in Figures \ref{SL_var} and \ref{F_var}.

The $B_{\rm hi}=500$~G data points in Figure \ref{Fig:transmission} correspond to the flat magnetic field case, where the inhomogeneity in $v_{\rm A}$ is small with $\lambda/L_{\rm A} < 0.5$. $\mathit{\Gamma}_{3}/\mathit{\Gamma}_{2}$ is $\approx 0.54$ for the three lower frequency measurements and $\approx 0.41$ for the two higher frequencies. The vertical error bars for all five frequencies nearly overlap. We attribute these minor differences to damping mechanisms that are most readily observable in a uniform plasma.  No such similar differences versus frequency were seen in our gradient-driven results, which show an almost monotonic reduction in $\mathit{\Gamma_{\rm 3}}/\mathit{\Gamma_{\rm 2}}$ with increasing $\lambda/L_{\rm A}$. For the gradient cases, the observed energy reduction relative to the non-gradient cases is substantial, with a decrease by a factor of $\approx 5$. Moreover, the monotonic nature of the decrease strongly suggests that the energy reduction is due to a gradient-driven effect.

\section{Analysis}\label{Sec:analysis}

In this section we first develop a model for the damping of shear Alfv{\'e}n waves in a uniform magnetic field in order to understand the reduction in wave energy for the homogeneous case. We then move on to explore the cause of the observed reduction in wave energy of waves propagating through a $v_{\rm A}$ gradient for the inhomogeneous case.   

\subsection{Reduction of wave energy in a uniform magnetic field}\label{Sec:damping_uniform}

 The flat field data in Figure \ref{Fig:transmission} show that shear Alfv{\'e}n waves damp in LAPD while propagating in a uniform 500 G magnetic field. Shear Alfv{\'e}n waves propagating in a uniform plasma are known to lose energy due to Landau damping and collisions \citep{cramer2011physics}.  Below we present numerical calculations quantifying the contribution of these two processes to the observed wave energy reduction. 
 
\subsubsection{Antenna model}\label{antenna_model}

The two dimensional structure of the wave magnetic field in Figure \ref{vector_2D} suggest that the ring antenna was in effect was driving two counter-propagating current channels along the axial magnetic field lines of LAPD. Field-aligned time-varying currents with frequencies below $f_{\rm ci}$ are known to radiate shear Alfv{\'e}n waves \citep{morales1994structure,gekelman1994experimental,morales1997structure}. Hence, we have modeled the antenna as two current sources driving field-aligned currents that are $180^{\circ}$ out of phase with one another.

More specifically, the ring antenna located at $x=y=z=0$ is modeled as two discs separated by a distance equal to the diameter of the ring, which is $\approx 9\rm \;cm$. The current density across the surface of each disc is assumed to have a Gaussian profile, $j_{0}\exp [-r^2/a^2 ]$, where $j_{0}$ is the amplitude of the surface current density, $r$ is radial the distance from the center of the disc, and $ a$ is a measure of the width of the current source. The wave magnetic field due to each current source lies in the azimuthal plane.  This azimuthal wave magnetic field, $b_{\phi}$, due to each disc is given by \citep{morales1997structure, vincena1999}
\begin{equation}
b_{\phi}=\frac{2 j_{0}\pi a^2}{\rm c} \int_{0}^{\infty}  \exp \left[ {-\frac{a^2 k_{\perp}^2}{4}} \right] J_{1}\left(k_{\perp}r \right) \exp\left[i k_{\parallel}\left(k_{\perp} \right)z\right] \mathrm{d}k_{\perp}. \label{Eq:azimuthal_B1}
\end{equation}   
\noindent
Here, $z$ is the axial distance from the antenna at which the wave magnetic field is calculated, $i = \sqrt{-1}$, and $J_{1}$ is the Bessel function of the first kind of order one, and $k_{\parallel}=k_{\rm r, \parallel} + i k_{\rm i, \parallel}$ is a  complex quantity where the real and imaginary parts are inversely proportional to the wavelength and damping length, respectively. Please note that $k_\parallel$ is a complex quantity only in the formulas mentioned here in Subsection \ref{Sec:damping_uniform}. In all other Sections and Subsections in this paper, $k_{\parallel}$ is a real quantity as defined in Section \ref{corona_lab}.

In order to simplify the calculation, we have normalized $b_{\phi}$ by the constant factor $2 j_{0}\pi a^2/c$ to obtain, 
\begin{equation}
b_{\phi, n}=\frac{b_{\phi}{c}} {  2 j_{0}\pi a^2}.\label{Eq:azimuthal_B}
\end{equation}
\noindent
The sole purpose of developing this model is to determine the damping length of the wave energy, which depends on the relative decrease of wave energy vs.\ the distance from the antenna. Our damping length results are not affected by this normalization.  

\subsubsection{Landau and collisional damping}

Landau damping is described by the warm plasma collisionless dispersion relation of shear Alfv{\'e}n waves derived from the linearzied Vlasov equation and Maxwell's equations \citep{swanson1989plasma}. This relation is given by \citep{stasiewicz2000small, lysak2008dispersion, thuecks2009tests}  
\begin{equation}
Z'\left(\xi\right) \left[\frac{v_{\rm A}^2}{2 v_{\rm te}^{2}} \frac{\left( 1 -\bar{\omega}^{2} \right)\mu_{\rm i} }{1-\Gamma_{0}(\mu_{\rm i} )} -\xi^2 \right]=k_{\perp}^2\delta^{2},
\label{Eq:landau_damping}
\end{equation}  
\noindent
where $\xi=\omega/\left(\sqrt{2}k_{\parallel}v_{\rm te}\right)$, $Z^{\prime}\left(\xi \right)=-2\{1+\xi Z\left( \xi \right) \}$ is the derivative of the plasma dispersion  function, $Z$,  \citep{fried1961plasma} with respect to $\xi$, $\mu_{\rm i}=k_{\perp}^{2} \rho_{\rm i}^{2}$ , $\rho_{\rm i} = {m_{\rm i}} v_{\rm ti}/qB_{0}$ is the ion gyroradius, $\Gamma_{0}\left( \mu_{\rm i} \right)=e^{-\mu_{\rm i}}I_{0}(\mu_{\rm i})$, and $I_{0}$ is the modified Bessel function of order zero.  

Collisional damping is modeled in the wave dispersion by including the Krook collision operator in the linearized Vlasov equation \citep{gross1951plasma, swanson1989plasma}.  The resulting dispersion relation is given by  \citep{gekelman1997laboratory, thuecks2009tests} 
\begin{equation}
Z'\left(\eta \right) \left(1 +i \frac{\nu_{\rm e}}{\omega}  \right) \left[\frac{v_{\rm A}^2}{2 v_{\rm te}^{2}} \frac{\left( 1 -\bar{\omega}^{2} \right)\mu_{\rm i}}{1-\Gamma_{0}(\mu_{\rm i})} -\xi^2 \right]=k_{\perp}^2\delta^{2},
\label{landaucoll_eq}
\end{equation}  
\noindent
where $\eta=\xi \left(1+i \nu_{\rm e}/\omega \right)$ and $\nu_{\rm e}$ is the collision frequency for electrons. The collision frequency, $\nu_{\rm e}=\nu_{\rm ei} + \nu_{\rm en}$, where $\nu_{\rm ei}$ and $\nu_{\rm en}$ are the electron-ion and electron-neutral collision frequency, respectively. From this dispersion relation we determine $k_{\parallel}$ as a function of $k_{\perp}$, which we then substitute into Equation (\ref{Eq:azimuthal_B}) to include the effect of Landau and collisional damping in the model. 

The electron-ion collision frequency is calculated in units of Hz using the expression \citep{Braginskii1965}
\begin{equation}
\nu_{\rm ei} = 2.9 \times 10^{-6} \frac{Z_{\rm ch}n {\rm ln \Lambda}}{T_{\rm e}^{3/2}}, \label{Eq:coll_freq}
\end{equation}  
\noindent
where $n$ is in $\rm cm^{-3}$ and $T_{\rm e}$ is in eV. The electron-neutral collision frequency is determined using the formula given by \citet{baille1981effective}. For our experimental parameters of $n=2.6\times10^{12} \;\rm cm^{-3}$, neutral pressure of $10^{-4}$ torr, and $T_{\rm e}=\rm 5 \; eV$ in a $\rm He^{+}$ plasma, $\nu_{\rm ei}$ is $\approx 7.5 \times 10^{6} \rm \; Hz$ and $\nu_{\rm en}$ is $\approx 4 \times 10^{5} \;\rm Hz$.

Equation (\ref{landaucoll_eq}) gives the dispersion relation of shear Alfv\'en wave in the presence of Landau and collisional damping. In order to determine the damping due to collisions only, we have used the collisional dispersion relation of shear Alfv\'en waves derived using the two fluid theory in the $\bar{\beta} \gg 1$ limit, \citep{vranjes2006unstable, gigliotti2009generation}
\begin{equation}
\omega^{2} - k_{\parallel}^2 v_{\rm A}^2 \left(1-\bar{\omega}^2 + k_{\perp}^2 \rho_{\rm s}^2 \right)+i \omega k_{\perp}^2 \delta^2 \nu_{e} = 0. 
\label{Eq:collisional_KAW}
\end{equation}  
\noindent

This two fluid dispersion relation can be solved algebraically, where $k_{\rm r, \parallel}$ and $k_{\rm i, \parallel}$ are given by
\begin{equation}
k_{\rm r, \parallel}= \frac{1}{\sqrt{2}} \frac{\omega}{v_{\rm A}} \left[\frac{\sqrt{1+k_{\perp}^4 \delta^4 \left( \nu_e /\omega \right)^2}+1}{1-\bar{\omega}^2 + k_{\perp}^2 \rho_{\rm s}^2}\right]^{1/2}, 
\label{Eq:kr_collisional_KAW}
\end{equation}  
\noindent

\begin{equation}
k_{\rm i, \parallel}= \frac{1}{\sqrt{2}} \frac{\omega}{v_{\rm A}} \left[ \frac{\sqrt{1+k_{\perp}^4 \delta^4 \left( \nu_e /\omega \right)^2}-1}{1-\bar{\omega}^2 + k_{\perp}^2 \rho_{\rm s}^2} \right]^{1/2}, 
\label{Eq:ki_collisional_KAW}
\end{equation}  
\noindent
respectively. Using Equations (\ref{Eq:kr_collisional_KAW}) and (\ref{Eq:ki_collisional_KAW}), we have determined $k_{\parallel}$ as a function of $k_{\perp}$, and substituted $k_{\parallel}(k_\perp)$ in Equation (\ref{Eq:azimuthal_B}) to include the effect of only collisional damping in the model. 

\subsubsection{Wave propagation model}

In order to compare the structure of the experimentally measured wave magnetic field with that predicted by the model we first calculated $b_{\phi}$ in cylindrical coordinates system due to each current source using Equation (\ref{Eq:azimuthal_B}). The value of $b_{\phi}$ for each source is then converted from the cylindrical coordinates to Cartesian coordinates as $b_{\phi}=b_{x} \hat{x} + b_{y} \hat{y}$. The total wave magnetic field produced by the ring antenna can then be modeled by a linear superposition of $b_{x} \hat{x}+ b_{y} \hat{y}$ produced by the two disc sources. We refer to this total wave magnetic field as $b_{\perp}$. 

\begin{figure}
	\centering
	\includegraphics[scale = 0.44]{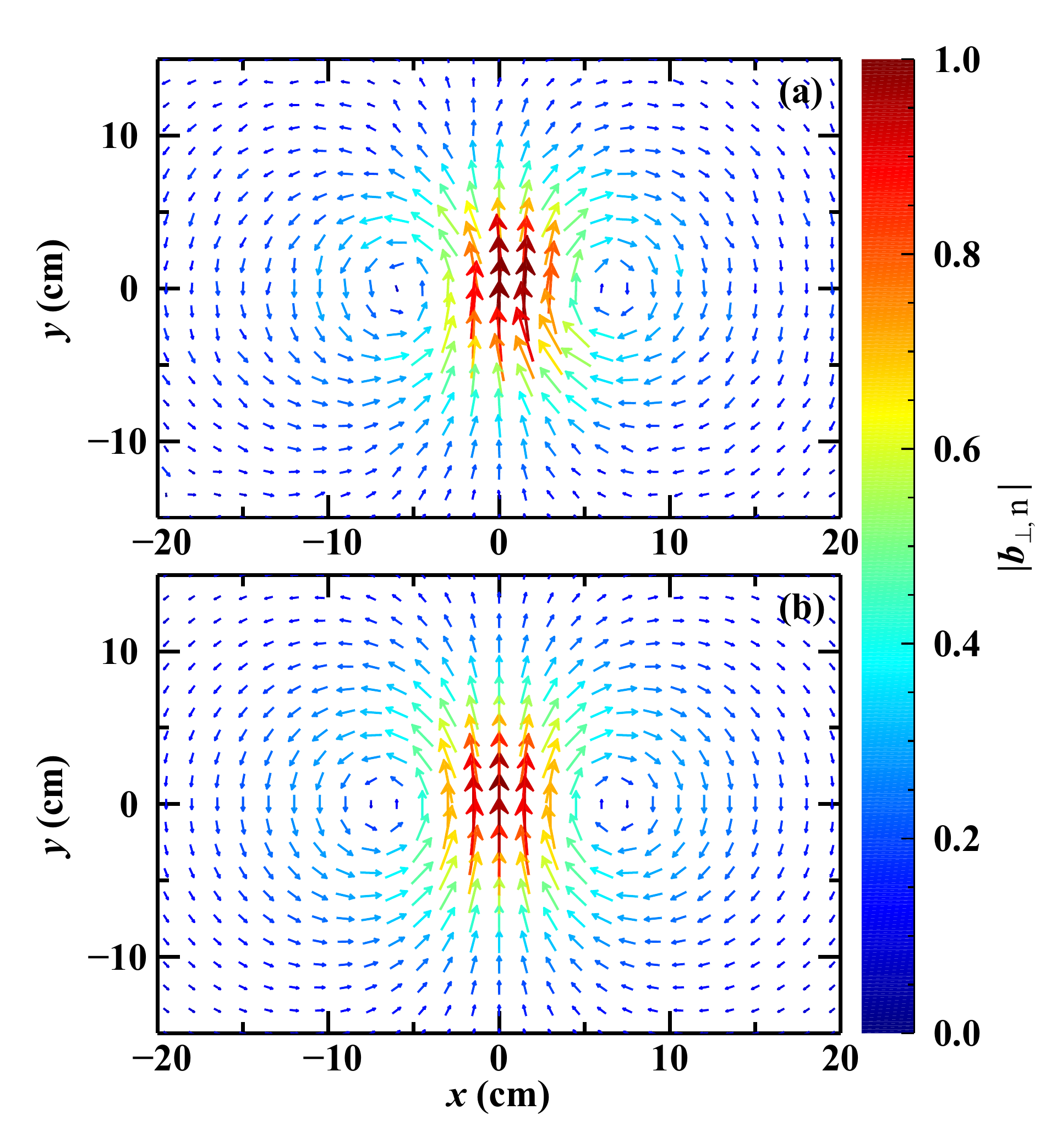}
	\caption{Comparison of the two dimensional wave structure of a 57.6 kHz shear Alfv{\'e}n wave  at $\rm z_{2}$ in a uniform 500 G magnetic field: (a) as measured by the B-dot  probe and (b) as predicted by the antenna and wave propagation model. See the text for additional details. 
		\label{Fig:2D_comparision}}
\end{figure}

The experimentally measured $b_{\perp}$ is compared in Figure \ref{Fig:2D_comparision} to the field calculated at $\rm z_{2}$ for $B_{0} = 500 \rm \; G$ and $f = 57.6\; \rm kHz$. The measured and calculated $b_{\perp}$, normalized by their maximum values, are denoted as $b_{\perp, \rm n}$. In the antenna model we set $a=0.25 \;\rm cm$, which is the thickness of the ring antenna. The average values of $n$ and $T_{\rm e}$ given in Table \ref{table:density_temp} were used for the numerical calculation.  Figure \ref{Fig:2D_comparision} shows that the wave structure predicted by the model using Equation (\ref{landaucoll_eq}). This structure is in excellent agreement with the measured wave magnetic field. The difference in the total wave energy, $\mathit{\Gamma}$, in the $xy$ plane calculated from the experimental and numerical data is typically $4\%$.

\begin{figure}
	\centering
	\includegraphics[scale = 0.44]{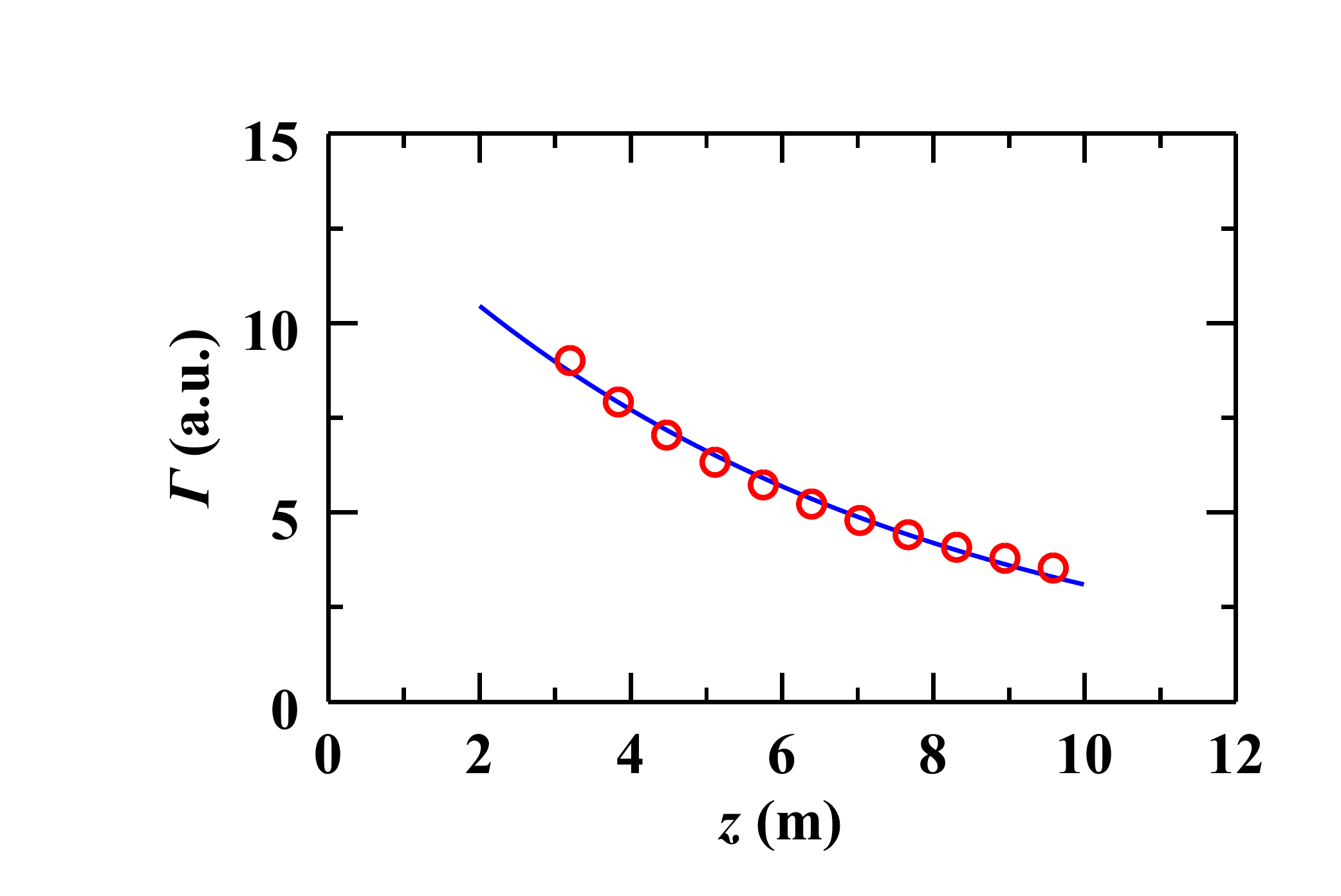}
	\caption{Damping of the modeled shear Alfv{\'e}n wave energy vs.\ distance along the LAPD axis. The wave energy values calculated using the model are given by red circles. The blue line is obtained by fitting a function of the form $A \exp[{-z/d}]$. 
		\label{Fig:57p6_damping}}
\end{figure}

The damping of wave energy in the model was determined by calculating Equation (\ref{Eq:azimuthal_B})  at multiple $z$ locations along the LAPD axis. Figure \ref{Fig:57p6_damping} presents the calculated damping of the shear Alfv{\'e}n wave shown in Figure \ref{Fig:2D_comparision}. The energy decay follows an exponential curve to a good approximation. The damping length, $d$, is obtained by fitting a function of the form $A \exp[{-z/d}]$ to the numerically calculated data, where $A$ is a constant. 

The reason that the energy damping curve of the shear Alfv{\'e}n wave approximately follows an exponential curve may be understood as follows. The wave magnetic field is obtained by inegrating over a number of $k_{\perp}$ as discussed by \citet{morales1997structure}. For a shear Alfv{\'e}n wave of a given frequency, different values of $k_{\perp}$ have different damping lengths as shown by \citet{gekelman1997laboratory}, \citet{kletzing2003measurements}, and \citet{lysak2008dispersion}. The cumulative effect of these multiple $k_{\perp}$ results in the energy decay being approximately exponential.

\begin{figure}
	\centering
	\includegraphics[scale = 0.41]{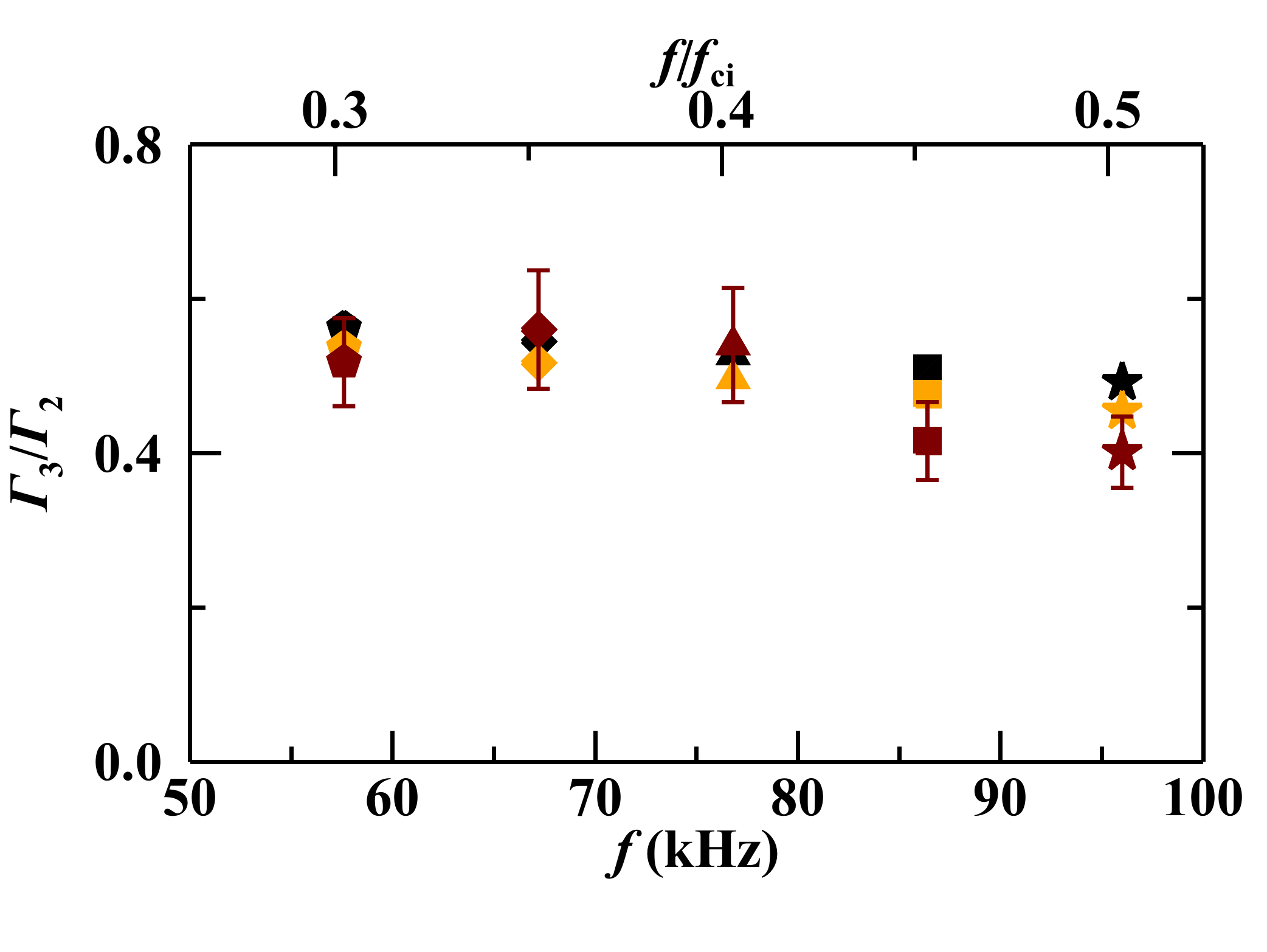}
	\caption{Comparison of the experimentally measured reduction in wave energy, $\mathit{\Gamma}_3/\mathit{\Gamma}_2$, between $\rm z_{2}$ and $\rm z_{3}$ and the theoretically calculated decrease in wave energy, using the antenna wave propagation model for $B_{\rm lo}=B_{\rm hi}=500$~G. The maroon symbols represent the experimental data. The black and orange symbols, respectively, represent the energy reduction due to collisional damping only and to the combined effects of Landau and collisional damping.   
		\label{Fig:damping_geometry}}
\end{figure}

The measured and modeled results for $\mathit{\Gamma}_3/\mathit{\Gamma}_2$ are shown in Figure \ref{Fig:damping_geometry}. The errors in the numerically calculated data were determined using a Monte Carlo method, by considering the uncertainties in $n$ and $T_{\rm e}$ given in Table \ref{table:density_temp}, and are on the order of the size of the plotted symbols. 

Comparing our observed damping for the flat 500~G case to our model results, we find that the wave energy reduction predicted by the model by considering both Landau and collisional damping is in good agreement with the experiment. The comparison shown in Figure \ref{Fig:damping_geometry} of the modeled results due to Landau and collisional damping, and only collisional damping shows that collisional damping is dominant. Landau damping while present, is very weak.

\subsection{Reduction of wave energy in the gradient}\label{Sec:wave_in_gradient}

\subsubsection{Wave reflection}\label{Sec:wave_reflection}

Light and other electromagnetic waves undergo reflection while propagating across a change in refractive index, corresponding  to a change in the propagation velocity of the wave. Similarly, according to both theoretical studies and numerical simulations, shear Alfv{\'e}n waves propagating through a strong longitudinal $v_{\rm A}$ gradient are predicted to undergo reflection \citep{moore1991alfven, musielak1992klein, perez2013direct}. 

According to the theory of \cite{musielak1992klein}, a shear  Alfv{\'e}n wave incident on a longitudinal $v_{\rm A}$ gradient is expected to undergo strong reflection when the frequency of the wave is less than the critical frequency $f_{\rm cr}$ given by 
\begin{equation}
f_{\rm cr}=\frac{1}{2}\sqrt{\left(v'_{\rm A}\right)^{2}+|2v_{\rm A}v''_{\rm A}|}. \label{critical_freq}
\end{equation} 
\noindent
Here, the double primes indicate the second spatial derivative. This expression was deduced for gradients in $n$ and $B_{0}$ in one dimension.

In the wave experiments described in Section \ref{SAW_propagation}, the shear Alfv{\'e}n waves passes through two gradients in $v_{\rm A}$ labeled as I and II in Figure \ref{exp_set_up}. The difference between these gradients is that $v_{\rm A}$ increases with distance in gradient I, while it decreases with distance in gradient II.  In order to constrain the role of reflected waves on the observed reduction in the transmitted wave energy vs.\ $\lambda/L_{\rm A}$, we performed several measurements to measure the magnitude of any reflected waves.  

\begin{figure}
	\centering
	\includegraphics[scale = 0.4]{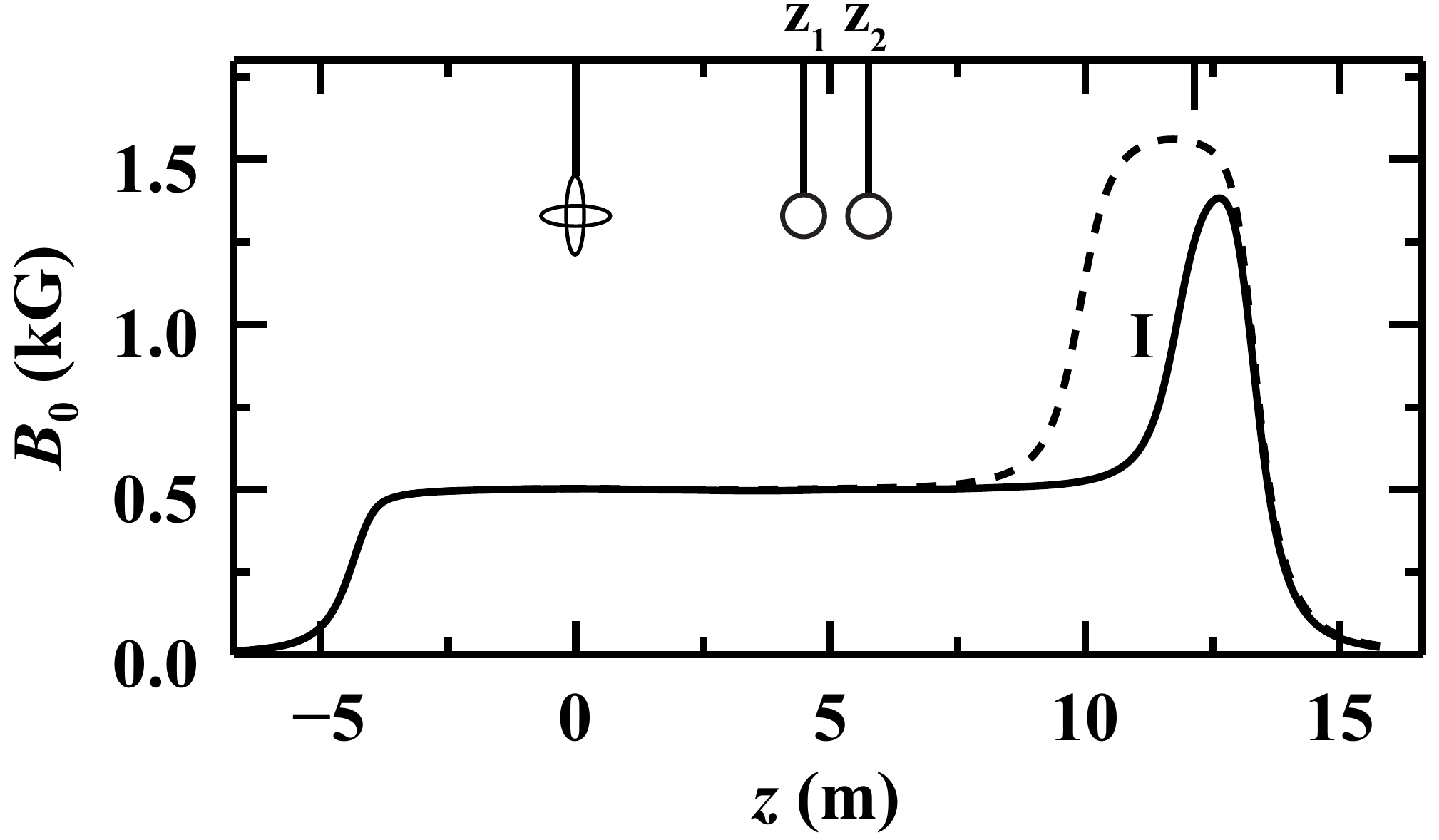}
	\caption{Magnetic field profile used for studying reflection of wave energy from the $v_{\rm A}$ gradient labeled as I. The locations $\rm z_{1}$ and $\rm z_{2}$ are the same as in Figure \ref{exp_set_up}.    
		\label{ref_B_profile}}
\end{figure}

The first two sets of wave-reflection experiments were carried out to search for reflection from gradient I. B-dot probes were positioned at $\rm z_{\rm 1}$ and $\rm z_{2}$ as indicated in Figure \ref{ref_B_profile}. Shear Alfv{\'e}n waves were excited by applying a sinusoidal wave train of two cycles to the antenna. This reduced the temporal length of the wave train compared to the previously excited ten-cycle wave train. The gradient was also moved to the far end of the machine, as shown in Figure \ref{ref_B_profile}. This ensured that the time required for a wave to traverse the distance from the B-dot probe at $\rm z_{1}$ to gradient I and return back to $\rm z_{1}$ was greater than twice the time period of the lowest wave frequency investigated. As a result, the incident wave and any reflected wave would be separated in time in the B-dot probe data. Lastly, the range of frequencies were selected to satisfy the  criteria: (a) the wave was predicted to be strongly reflected by theory and (b) there was an overlap in the values of $\lambda /L_{\rm A}$  with the inhomogeneity observed in coronal holes of $\lambda/L_{\rm A} \gtrsim 4.5$.

The magnetic field profile used in the first set of wave-reflection experiment is shown by the solid curve in Figure \ref{ref_B_profile}. This curve was obtained by increasing $B_{\rm hi}$ to $\approx 1382\; \rm G$. The value of $f_{\rm cr}$ varied axially vs.\ the magnetic field variation and reached a maximum value of $\approx 724$ kHz at $ z=\rm12.14 \;m$ within gradient I. In order to satisfy the theoretical criteria for strong reflection we excited wave frequencies below 724 kHz.

\begin{figure}
	\centering
	\includegraphics[scale = 0.4]{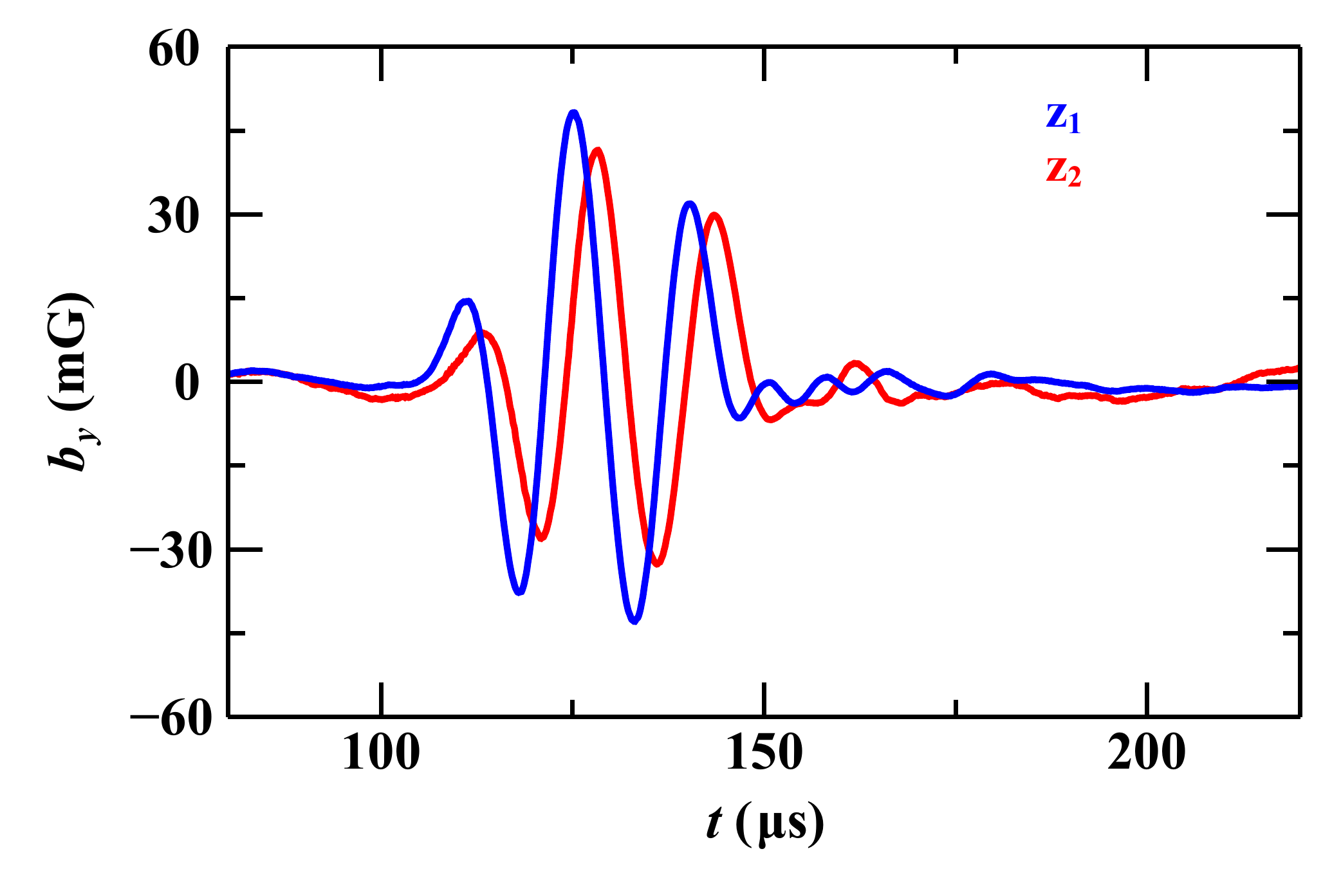}
	\caption{Time variation of $b_{y}$ as measured on axis of LAPD at $\rm z_{1}$ and $\rm z_{2}$ for $f = $ 65 kHz.  The magnetic field profile corresponding to this wave data is given by the solid line in Figure \ref{ref_B_profile}.       
		\label{ref_grad1}}
\end{figure}

Figure \ref{ref_grad1} shows the $y$ component of the wave magnetic field detected for $f=65 \rm \; kHz$ (i.e.,  $f/f_{\rm cr}=0.09$). A well formed two-cycle incident wave was detected by probes at $\rm z_{1}$  and $\rm z_{2}$ between $\approx$ 112 and 148 $\mu\rm s$. For this wave $\lambda / L_{\rm A}$ was $ \approx 5.6$. If the shear Alfv{\'e}n wave were strongly reflected by gradient I, then the reflected wave would reach $\rm z_{1}$ at   $39\;\rm \mu s$ after the incident wave has passed it. Hence, a reflected wave is predicted to be observed in Figure \ref{ref_grad1} between $\approx $ 151 to 187 $\rm \mu s$. However, we do not observe any reflected wave in this time window. The wave signal at $\rm z_{1}$  always leads the signal at $\rm z_{2}$ implying that the B-dot probes did not detect any waves reflected by gradient I. We also find no detectable reflected wave in the $b_{x}$ and $b_{z}$ directions

There are some small amplitude fluctuations trailing the applied two-cycle wave train, but these features have a frequency twice that of the applied waveform. We believe that these are excited by the second harmonic present in the antenna signal and are unrelated to reflection from gradient I.

We have also considered the possibility that waves did not reflect back exactly along the axis, but found no evidence for reflected waves at any location in the LAPD cross section. The B-dot probes were scanned through a cross section in LAPD and the results were always similar to those shown in Figure \ref{ref_grad1}. 

We repeated all of these measurements while increasing $f$ to 190~kHz in steps of 5 kHz. This caused $f/f_{\rm cr}$ to increase to 0.3. As before, we did not observe a detectable reflected wave at any of the measured frequencies.

We then carried out a second set of experiments to study the possible effects of tunneling through the high $v_{\rm A}$ region. For the above wave-reflection studies, $\lambda$ was greater than the width, $w$, of the high field region. For example, for $f/f_{\rm cr} = 0.1$, $ \lambda/w$ was $ \approx 2.1$. To rule out the possibility that the shear Alfv{\'e}n wave could be tunneling through the high $v_{\rm A}$ region, instead of undergoing reflection, we  moved gradient I closer to the antenna, as represented by the dashed line in Figure \ref{ref_B_profile}. This resulted in $\lambda /w  \approx 1.1$. We then repeated the wave-reflection experiments described above and again did not observe any detectable reflected waves.

The lack of an observable reflected wave from gradient I may imply that the amplitude of the reflected wave is too weak to be detected. For example, using the measured initial wave amplitude at z$_1$ and taking into account Landau and collisional damping as the wave propagates from z$_1$ to gradient I and back, we estimate that if there were 100\% reflection at gradient I, then we would measure a reflected wave signal at z$_1$ with an 11~mG amplitude.  This is much larger than the $\approx 1.5$~mG fluctuations in Figure \ref{ref_grad1} trailing the applied two-cycle wave train and should be readily observable.  The lack of an observed reflected signal indicates that the efficiency of any reflection by the gradient is much less than 100\%.  Taking 3~mG as a reasonable detectable level over the 1.5~mG fluctuations trailing the applied wave train, and taking into account Landau and collisional damping between z$_1$ and gradient I and back to z$_1$, we can put an upper limit on the reflected wave energy of $\approx 7.4 \%$ for $\lambda / L_{\rm A} \approx 5.6$, and $f/f_{\rm cr}=0.09$. This reflectance is too small to account for the observed wave energy loss.

\begin{figure}
	\centering
	\includegraphics[scale = 0.4]{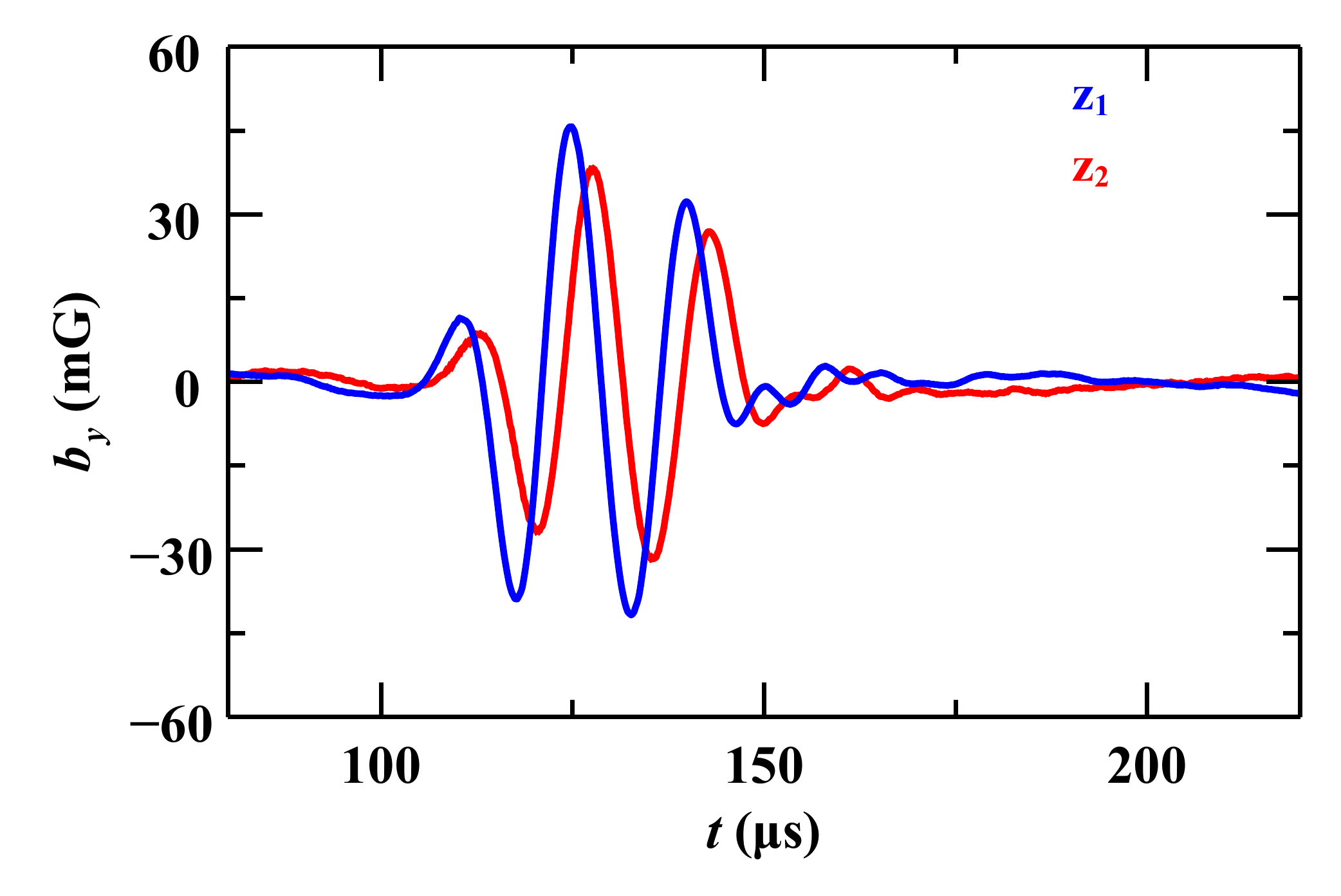}
	\caption{Same as Figure 13 but for $B_{\rm lo} = B_{\rm hi} = 500$~G.       
		\label{500_grad2}}
\end{figure}

Finaly, in the third set of wave-reflection experiments, we investigated the effects of gradient II. For this we set $B_{\rm lo} = B_{\rm hi} = 500$~G, the flat field case shown in Figure \ref{exp_set_up}. Here, $f_{\rm cr}$ had a maximum value of $\approx 364 \rm \; kHz$ within gradient II at $z = 13\;\rm m$. As before shear Alfv{\'e}n waves were excited by applying a sinusoidal wave train comprising of two cycles to the antenna located at $z=0$. The corresponding variation of $b_{ y}$ at $\rm z_{1}$ and $\rm z_{2}$ for $f=65\; \rm kHz$ is shown in Figure \ref{500_grad2}. Similar to the results of the first two sets of wave-reflection measurements, the phase of the wave signal at $\rm z_{1}$ always leads the wave signal at $\rm z_{2}$, indicating that the B-dot probes did not detect any reflection from the gradient.

\subsubsection{Mode coupling}

The energy of a shear Alfv{\'e}n wave traveling through a $v_{\rm A}$ gradient may decrease if a part of the wave energy is converted into another mode. Inhomogeneity in the magnetic field can enable the propagation of compressible surface magnetoacoustic waves and incompressible surface Alfv\'en waves \citep{roberts1981wave}. A gradient in the magnetic field may convert some of the shear Alfv\'en wave energy to a slow wave \citep{southwood1985curvature}. A fast wave may get excited. All of these modes induce a parallel perturbation, which we have tried to detect using B-dot probes. The ratio  $b_{\parallel}/b_{\perp}$ was measured  before, within, and after the gradient. Mode conversion into these modes would produce an amplification of $b_{\parallel}$; but we did not detect any $b_{\parallel}$ above the noise level of $\approx$ 0.5~mG. This implies that mode conversion is unlikely. 

\begin{figure}
	\centering
	\includegraphics[scale = 0.24]{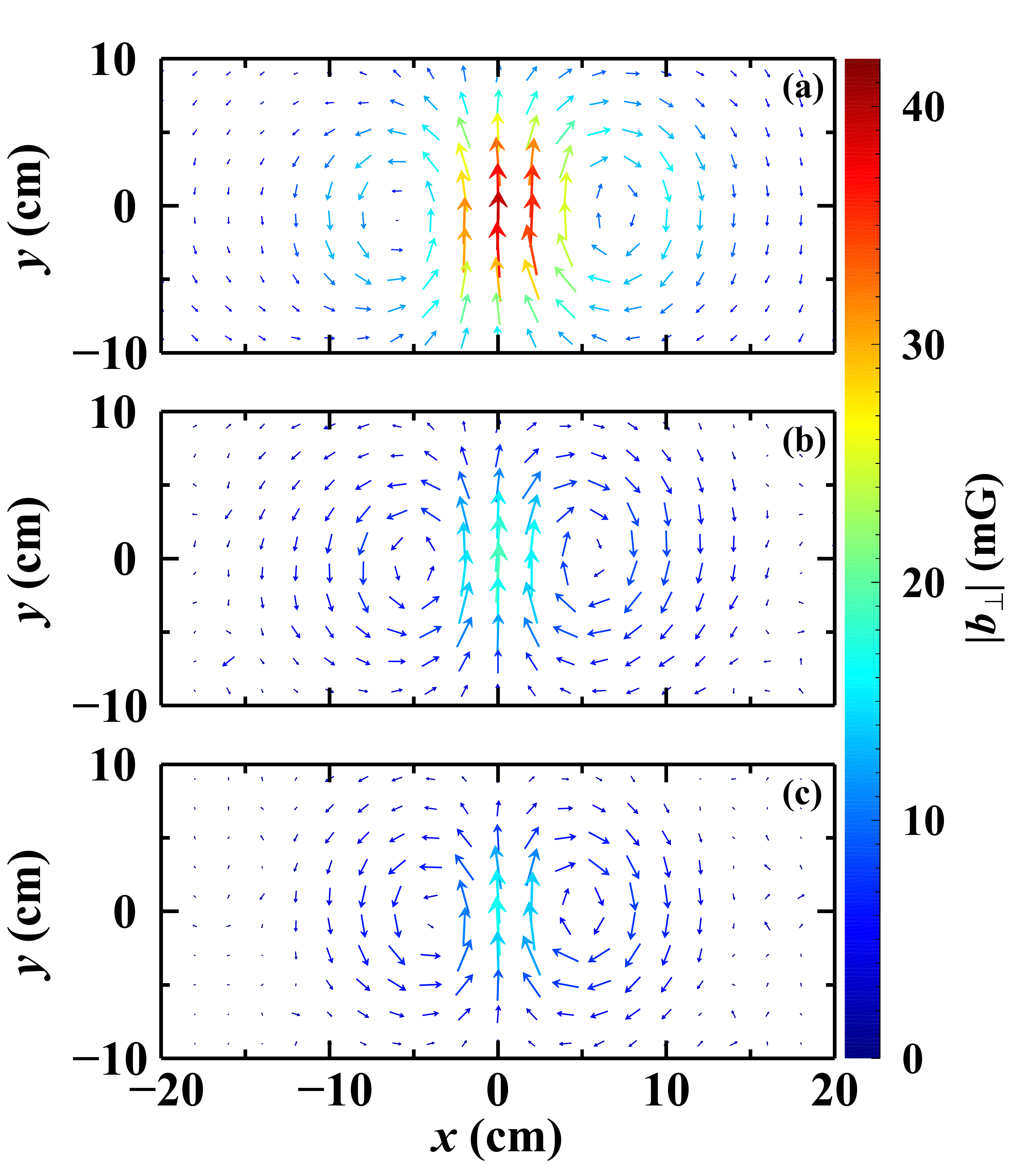}
	\caption{Wave magnetic field in the $xy$ cross-section of LAPD (a) before the gradient at $z=z_{2}$, (b) within the gradient at $z =$ 8~m, and (c) after the gradient at $z=z_{3}$. The direction of the arrows represent that of the wave magnetic field and the colors give the magnitude of the field using the color bar shown. The arrow lengths are normalized by the maximum value of the magnetic field in each panel. For these measurements $B_{\rm lo}$ was 500~G, $B_{\rm hi}$ was 1600~G, and the wave frequency was 76.8 kHz. Note that the $x$ and $y$ spacing of the data points here is 2~cm as opposed to 1.5~cm in Figure \ref{vector_2D}.             
		\label{fig:bef_in_aft}}
\end{figure}

In order to further confirm that the only mode propagating in the gradient is a shear Alfv\'en wave, we have measured the wave magnetic fields before, within, and after the gradient. Figures \ref{fig:bef_in_aft} (a), (b), and (c) show the structure of the wave before, within and after the magnetic field gradient, respectively. Two well formed current channels are observed with the separation between the current channels decreasing in the gradient and on the high field side, as is expected for shear Alfv\'en waves propagating along converging magnetic field lines. Considering that we have not detected any wave other than the shear Alfv\'en wave in the gradient, and that the structure of the shear Alfv\'en wave within the gradient is consistent with that before and after the gradient, energy reduction due to mode conversion is unlikely.

\subsubsection{Nonlinear effects and transit time damping}\label{sec:nonlinear_effects}

Large amplitude shear Alfv{\'e}n waves can lose energy due to nonlinear effects. The experiments reported in this paper were carried out using very low amplitude waves of $b/B_{0} \lesssim 8 \times 10^{-5}$. Nonlinear effect associated with shear Alfv{\'e}n waves, such as parametric instability, have been found to occur only for relatively large amplitude waves. For example, \cite{dorfman2016observation} reported the threshold for observation of parametric instability of shear Alfv{\'e}n waves to be $b/B_{0} \geq 2 \times 10^{-3}$. This is over 10 times greater than our wave amplitude. Low amplitude shear Alfv{\'e}n waves also exhibit nonlinear effects in a narrow band of frequencies around the ion cyclotron frequency due to ion cyclotron resonance. As the range of wave frequencies excited here is $\le f_{\rm ci}/2$, nonlinear effects due to ion cyclotron resonance are expected to be absent.    

A shear Alfv\'en wave may exert a mirror force on the electrons and ions, and contribute to additional damping of the wave. This damping mechanism is called transit-time damping. For uniform plasmas with $v^2_{\rm ti} \ll v_{\rm A}^2$, and $T_{\rm i} < T_{e}$,  the mirror force experienced by an electron is greater than that for an ion, and  is given by $|F_{\rm Me}| \sim \left(m_{e} v_{te}^2/2B_{0}\right) k_{\parallel} b_{\parallel}$ \citep{hollweg1999kinetic}. In order to estimate the relative importance of transit-time damping with respect to collisional damping, we have compared $|F_{\rm Me}|$ with the frictional force experienced by an electron, $|F_{\rm fric,e}|\sim m_{e}v_{\rm te } \nu_{e}$ \citep{swanson1989plasma}. For our experimental parameters of $n=2.8\times10^{12} \;\rm cm^{-3}$, neutral pressure of $10^{-4}$ torr, $T_{\rm e}=\rm 4.9 \; eV$, maximum value for $b_{\parallel}$ of $ 0.5~\rm mG$, $B_{0} = $ 500~G, and $k_{\parallel} = $ 0.014~m, the estimated $|F_{\rm Me}|$/$|F_{\rm fric, e}|$ is $\sim  8\times 10^{-8}$. This is extremely small, mainly because $b_{\parallel}/B_{0} < 10^{-6}$ in the gradient. Hence, the transit-time damping due to mirror force is inconsequential when compared to collisional damping and cannot account for the reduction in wave energy.

\section{Discussion and summary}\label{dis_sum}

We have studied the reduction in energy of shear Alfv{\'e}n waves propagating through $v_{\rm A}$ gradients in a laboratory experiment under conditions scaled to match solar coronal holes. We have experimentally established that $\lambda/L_{\rm A}$ is the independent parameter that describes the decrease in energy of shear Alfv\'en waves passing through $v_{\rm A}$ gradients. For values of $\lambda/L_{\rm A}$ similar to those in coronal holes, the waves are observed to lose energy by a factor of $\approx 5$ more than they do when propagating through a plasma without a gradient, where the energy reduction is by a factor of $\approx 2$. 

In the absence of a magnetic field gradient, we have used a model to show that the wave energy reduction is caused by collisional and Landau damping. Collisions are found to dominate to the wave damping, while the contribution of Landau damping is small. 
 
The cause of the additional damping in the presence of a gradient is unknown. We have constrained the cause of this energy reduction in the gradient by ruling out wave reflection, mode coupling,  non-linear effects, and transit-time damping. Landau and collisional damping may reduce the energy of the shear Alfv\'en wave in the gradient and deposit the wave energy in the plasma. However, a detailed theoretical analysis to accurately determine their contribution using plasma kinetic theory in the non-WKB regime relevant to our experiments is beyond the scope of this paper.     

Since the most probable mechanisms that can reduce the energy of the incident shear Alfv\'en wave without transferring the energy to the plasma do not account for the observed energy reduction, and as the total energy must be conserved, it is likely that the waves deposit their energy in the plasma, thereby contributing to plasma heating or generating a bulk flow. If the total energy lost by the wave contributed to electron heating, then the maximum increase in $T_{\rm e}$ would be $\sim 23\;\rm \mu eV$. Unfortunately this is too small of an increment to measure with a Langmuir probe. If the total wave energy gave rise to a bulk flow, then the flow velocity would be $v_{\rm f} \sim 0.003\; c_{\rm s}$, which is too weak to be detected with a Mach probe.  The reason for the expected small rise in $T_{\rm e}$ or weak $v_{\rm f}$ is due to the low amplitude of the shear Alfv{\'e}n wave. The energy lost by the shear Alfv{\'e}n wave is only $\sim 55\;\rm \mu J$. This is a tiny fraction of the $\sim 1\rm\; J$ thermal energy of the plasma in the longitudinal gradient. In future experiments we hope to determine the location where the wave energy is being deposited. In order to detect $\Delta T_{\rm e}$ or $v_{\rm f}$, we plan to excite large amplitude shear Alfv{\'e}n waves and carry out simultaneous $T_{\rm e}$ and $v_{\rm f}$  measurements as the wave propagates through the gradient.

\section*{Acknowledgment} 
The authors thank W. Gekelman for stimulating discussions. This material is based, in part, upon work supported by the U.S. Department of Energy, Office of Science, Office of Fusion Energy Science under Award Number DE-SC-0016602. The experiments were performed at the Basic Plasma Science Facility (BaPSF) which is supported by the DOE and NSF, with major facility instrumentation developed via an NSF award AGS $-\;9724366$.




\end{document}